\documentclass[%
 reprint,
 amsmath,amssymb,
 aps,
 reprint,
pra,
nofootinbib
]{revtex4-1}

\usepackage{ulem}
\usepackage[dvipsnames]{xcolor}

\usepackage{graphicx}
\usepackage{dcolumn}
\usepackage{bm}
\usepackage{hyperref}  
\hypersetup{colorlinks = true, linkcolor = [rgb]{0.19411,0.51882,0.667058}, urlcolor = [rgb]{0.125490,0.29542,0.1647058}, citecolor = [rgb]{0.75882,0.37411,0.14117}}

\usepackage{setspace}

\usepackage{amssymb}

\usepackage{braket}

\graphicspath{{../images/}}
\usepackage[toc,page]{appendix}

\newcommand{\nn}{\nonumber    \\}

\usepackage{braket}

\def\>{\rangle}
\def\<{\langle}

\usepackage[bottom]{footmisc}

\begin{document}
\title{{Quantum-limited estimation of range and velocity}}

\author{Zixin Huang}
\author{Cosmo Lupo}
\author{Pieter Kok}

\affiliation{Department of Physics and Astronomy, The University of Sheffield, Sheffield, S3 7RH, United Kingdom.}

\begin{abstract}
The energy-time uncertainty relation puts a fundamental limit on the precision of radars and lidars for the estimation of range and velocity. The precision in the estimation of the range (through the time of arrival) and the velocity (through Doppler frequency shifts) of a target are inversely related to each other, and dictated by the bandwidth of the signal.
Here we use the theoretical toolbox of multi-parameter quantum metrology to determine the ultimate precision of the simultaneous estimation of range and velocity.
We consider the case of a single target as well as a pair of closely separated targets. 
In the latter case, we focus on the relative position and velocity.
We show that the trade-off between the estimation precision of position and velocity is relaxed for entangled probe states, and is completely lifted in the limit of infinite entanglement. 
In the regime where the two targets are close to each other, the 
relative position and velocity can be estimated nearly optimally and jointly, even without entanglement, using the measurements determined by the symmetric logarithmic derivatives.

\end{abstract}
\date{\today}
 
\maketitle

\section{Introduction}
Quantum metrology \cite{giovannetti2004quantum,PhysRevLett.96.010401} and quantum imaging \cite{kolobov2007quantum} aim at exploiting physical resources such as quantum coherence and entanglement to achieve precision measurements and image resolution beyond those that are allowed by classical physics. 
A number of applications harness quantum correlations in the energy-time degrees of freedom \cite{PhysRevLett.62.2205}, including lithography \cite{PhysRevLett.87.013602}, quantum-enhanced positioning \cite{giovannetti2001quantum}, quantum illumination \cite{lloyd2008enhanced,PhysRevLett.101.253601,PhysRevLett.114.080503}, phase estimation 
\cite{PhysRevA.94.012101, PhysRevA.101.012124,PhysRevA.97.032333}, and ghost imaging \cite{shapiro2012physics}. 
The ability to accurately measure the temporal and spectral properties of light has led to significant developments in chemical spectroscopy \cite{zewail2000femtochemistry}, ranging \cite{mccarthy2009long,mccarthy2013kilometer}, clock synchronization \cite{RevModPhys.75.325}, continuous-variable superdense coding \cite{liu2020joint}, and quantum key distribution \cite{nunn2013large,PhysRevLett.112.120506}.

In radars and lidars, electromagnetic pulses are sent to interrogate a region of interest, and the back-reflected signals are collected and examined. 
The standard technique to resolve the target's longitudinal position is based on a measurement of the time of flight associated with a round trip. 
Furthermore, the target's radial velocity can be deduced by examining the Doppler-frequency shift of the returned signals.
In this paper, we analyse a model where faint pulses, containing at most one photon, are sent to interrogate a region of space that may contain one or two targets. 
The photons are prepared in wave packets with central frequency $\omega$. 
Our results are applicable to both lidar and radar, although the precision of radar is 
not quantum limited \cite{komissarov2019partially}.

If a pulse encounters a single target at distance $x$ that moves with relative (non-relativistic) radial velocity $v$, then the back-reflected photon (in a lossless scenario) will return after a time delay $\tau$, with its central frequency shifted to $\omega' = \omega + \delta\nu$. 
The range $x$ and velocity $v$ of the target can thus be estimated as 
$x \simeq c \tau / 2$, and 
$v \simeq c ( \delta\nu / \omega ) /2$, where $c$ is the speed of light.
If instead there are two close targets in the region of interest, with radial position $x_1$, $x_2$ and velocity $v_1$, $v_2$, 
a measurement of the time of arrival and the frequency shift allows us to estimate the central position $(x_1+x_2)/2$ and velocity $(v_1+v_2)/2$, as well as the relative position $x_1-x_2$ and relative velocity $v_1-v_2$.

The precision in the estimation of the time of arrival and of the signal frequency is \cite{skolnik1960theoretical,curry2005radar,giovannetti2001quantum}
\begin{align}
 \delta t \simeq \frac{\sigma_t}{ \sqrt{\nu n}} \, , \qquad 
 \delta \omega \simeq \frac{\sigma_\omega}{ \sqrt{\nu n}} \, ,
\end{align}
where $\sigma_\omega (\sigma_t)$ is the frequency (time) width of the signal, $N$ is the number of pulses used, and $n$ is the number of photons in each pulse. 
The Arthurs-Kelly uncertainty relation \cite{arthurs1965bstj} expresses the fundamental precision limit on the estimation of these parameters:
\begin{align}\label{AK-ineq}
 \sigma_t \sigma_\omega \geqslant 1/2 \, .
\end{align}
Thus for non-entangled photons, radar and lidar systems are subject to a fundamental trade-off in their ability to resolve the target's range and velocity. 

It is known that uncertainty relations may weaken in the presence of entanglement \cite{PhysRevLett.120.053601,Berta2010,PhysRevA.96.040304}, a phenomenon that could be exploited to boost the precision of quantum-limited range and velocity detection.
We use the toolbox of quantum information theory, in particular multi-parameter quantum metrology, to assess the ultimate precision of quantum-limited radars and lidars, with or without the assistance of entanglement. 
We consider the regime of faint pulses with at most one photon each, modelled using a Gaussian envelope. Within this model, we study the problem of jointly estimating the position and velocity of a target, as well as the relative position and velocity of two close targets.

Position and velocity estimation translates into time and frequency estimation, which has been considered before.
The estimation of time and frequency shifts following the detection of a single target was previously studied by Zhuang, Zhang, and Shapiro in the limit of very large entanglement \cite{PhysRevA.96.040304}.
Here, we consider time and frequency estimation for general probe states with arbitrary entanglement quantified by a continuous parameter $\kappa$.
The estimation of the relative time and frequency for two pulses was considered by Silberhorn and collaborators \cite{PhysRevLett.121.090501,ansari2020achieving}, but they did not consider the use of entanglement and the simultaneous estimation of these parameters.

The structure of the paper is as follows.
In Sec.~\ref{sec:tools} we briefly review the tools for multi-parameter quantum parameter estimation.
In Sec.~\ref{sec:model} we introduce our model.
Section~\ref{Sec:1target} presents the ultimate limit in the estimation of position and velocity of a target.
In Sec.~\ref{Sec:2targets} we determine the ultimate limit in the estimation of the relative position and velocity of two target, and an optimal measurement strategy (for the estimation of the relative position) is presented in Sec.~\ref{sec:measurement}.

\section{Theoretical toolbox}\label{sec:tools}

\begin{figure}[t!]
\includegraphics[trim = 0cm 0.5cm 0cm 0cm, clip, width=0.95\linewidth]{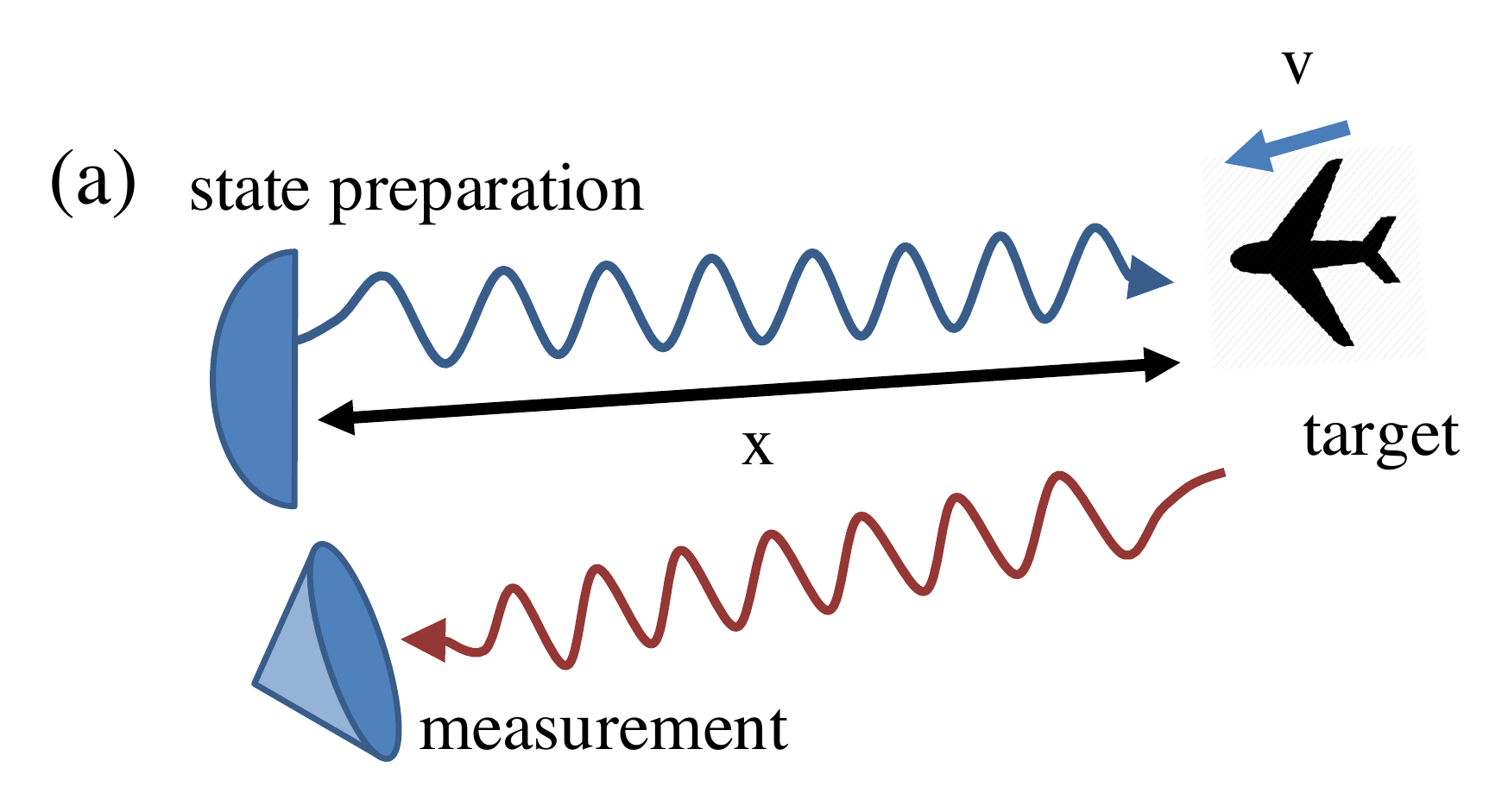} 
\includegraphics[trim = 0cm 0.5cm 0cm 0cm, clip, width=0.95\linewidth]{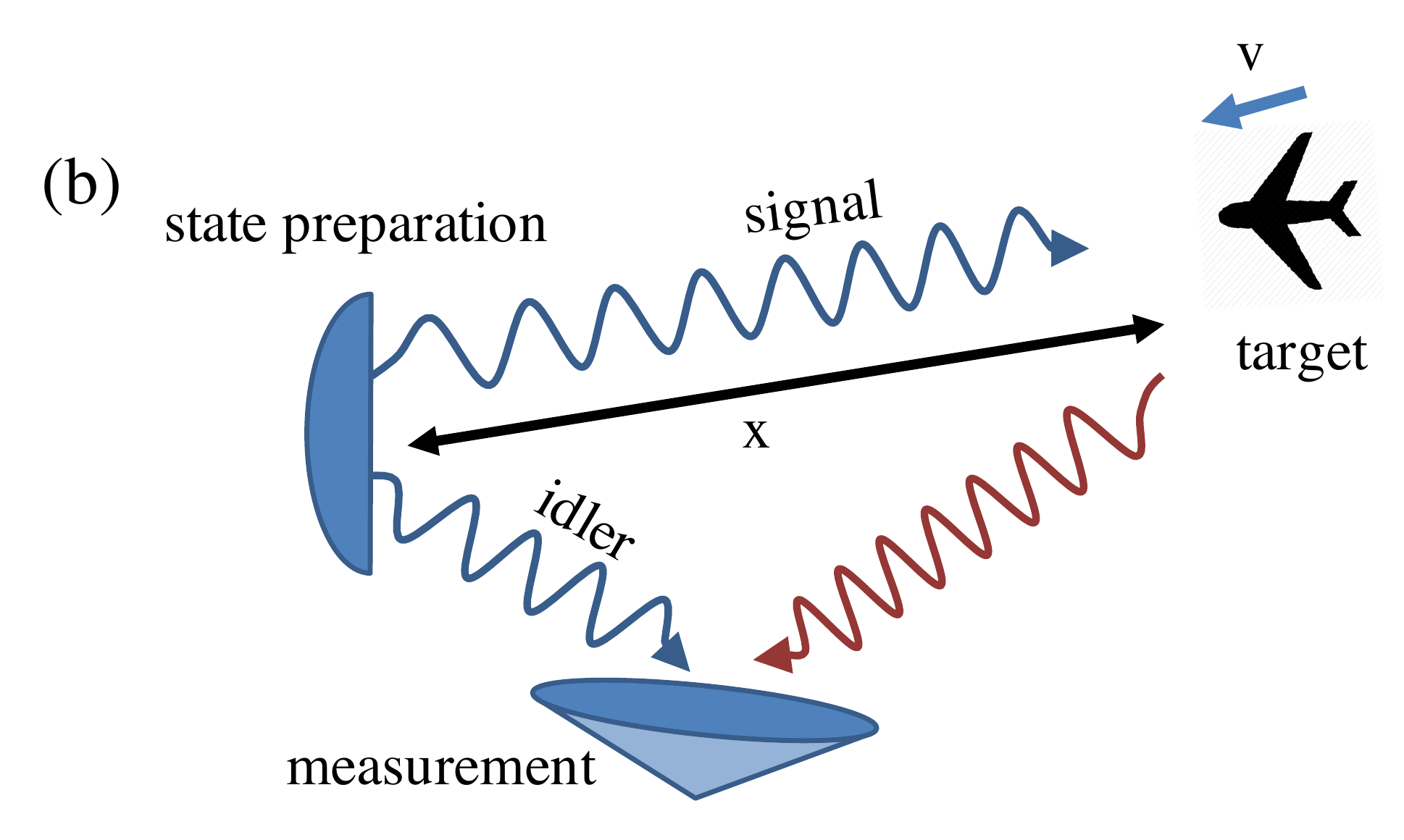}
 \caption{\label{f:scheme} Radar and lidar systems for measuring the velocity and range, using (a) separable and (b) entangled states.}
\end{figure}

A quantum parameter estimation routine typically consists of three stages, followed by a classical data processing step. 
This is shown schematically in Fig.~\ref{f:scheme}(a).
First, a quantum system is prepared in a known quantum state.
Second, the quantum system is used to probe a target system that we want to investigate.
Third, the probe is measured after interaction with the target.
In non-adaptive estimation strategies, the above is repeated $N$ times.
Finally, the raw data collected is processed to extract a best estimate for the parameters of interest.
An entanglement-assisted strategy refers to the scenario where the probe is initially entangled with an auxiliary system. The latter does not interact with the target, but it is jointly measured with the probe. This is shown in Fig.~\ref{f:scheme}(b). 

The ultimate precision in the estimation is given by the quantum Cram\'er-Rao (QCR) bound \cite{caves,caves1}.
For the estimation of the parameter $\lambda$ encoded onto a quantum state $\rho_\lambda$, this is a lower bound on the variance $\Delta\hat\lambda^2 = \langle \hat\lambda^2 \rangle - \langle \hat\lambda \rangle^2$ of any unbiased estimator $\hat\lambda$.
For unbiased estimators, the QCR bound establishes that 
\begin{align} \label{eq:var}
 \Delta \hat \lambda ^2 \geqslant \frac{1}{N} \frac{1}{J(\rho_\lambda)} \, ,
\end{align}
where $N$ is the number of probe systems used, and $J$ is the quantum Fisher information (QFI) associated with the global state $\rho_\lambda$ of the probes. 
The latter is defined as
\begin{eqnarray}
 J(\rho_\lambda)= \text{Tr}\left( L_\lambda^2 \rho_\lambda \right) \, ,
\end{eqnarray}
where $L_\lambda$ is the Symmetric Logarithmic Derivative (SLD) associated with the parameter $\lambda$ \cite{paris2009quantum}.
If the state $\rho_\lambda$ lives in a Hilbert space of dimensions $d$, consider a set of basis vectors $|e_1\rangle$, $|e_2\rangle$, $\dots$, $|e_d\rangle$ in which $\rho_\lambda$ is diagonal:
\begin{align}
 \rho_\lambda = \sum_n p_n |e_n\rangle \langle e_n| \, .
\end{align}
The SLD is then given by
\begin{align}
 L_\lambda = 2 \sum_{n,m : p_n+p_m \neq 0} \frac{\braket{e_m|\partial_\lambda \rho |e_n}}{p_n + p_m} \ket{e_m}\bra{e_n} \, ,
\end{align}
with $\partial_\lambda \rho = \frac{\partial \rho_\lambda}{\partial \lambda}$.
The QCR bound is asymptotically saturated, that is, there exists a measurement strategy and an unbiased estimator such that Eq.\ \eqref{eq:var} is tight in the limit that $N\rightarrow \infty$ \cite{PhysRevLett.110.240405}.
The SLD directly determines an optimal measurement, which is a projective measurement in the eigenvectors of the SLD operator.

When the quantum state $\rho_{\bm{\lambda}}$ carries information about multiple parameters, $\bm{\lambda} = \lambda_1, \dots, \lambda_K$, the statistical error in their estimation is expressed by the covariance matrix of the estimators $\bm{\hat\lambda} = \hat\lambda_1, \dots, \hat\lambda_K$,
\begin{align}
 \text{Cov}[\bm{\hat\lambda}]_{ij} = \braket{\hat\lambda_i \hat\lambda_j} - \braket{\hat\lambda_i } \braket{\hat\lambda_j} \, .
\end{align}
The multi-parameter QCR bound establishes the fundamental lower bound on the covariance matrix of any set of unbiased estimators. 
This is expressed as a matrix inequality
\begin{align}
 \text{Cov}[\bm{\hat\lambda}] \geqslant \frac{1}{N} J(\bm{\lambda})^{-1} \, ,
\end{align}
where $J(\bm{\lambda})$ is the QFI matrix, defined as
\begin{align}
J(\bm{\lambda})_{ij} = \frac{1}{2} \text{Tr}\left( \rho_{\bm{\lambda}} \{ L_{\lambda_i} , L_{\lambda_j} \} 
\right) \, .
\end{align}
Unlike the single-parameter case, there might not exist a single measurement that allows us to jointly estimate $K>1$ parameters simultaneously and optimally. 
This means that the multi-parameter QCR bound is not always achievable. 
A sufficient condition for the joint and optimal estimation is that the SLD operators commute. A weaker condition, which is necessary and sufficient, is %
\begin{align}\label{condition}
\text{Tr}\left( \rho_{\bm{\lambda}} [ L_{\lambda_i}, L_{\lambda_j} ] \right) = 0 \, .
\end{align}
If this condition holds, then there exists a single measurement and a set of $K$ estimators that saturate the multi-parameter QCR bound in the limit that $N \to \infty$ \cite{PhysRevA.94.052108,kay1993fundamentals}.

When the condition in Eq.~(\ref{condition}) is not met, there exists a tighter bound based on the so-called Right Logarithmic Derivative (RLD).
In their analysis, Zhuang \textit{et al}.\ employed this bound  \cite{PhysRevA.96.040304}, the particular form of which was proved by Fujiwara \cite{Fujiwara}. While the RLD bound is tighter, the RLD operator does not directly relate to a measurement operator.
Here, we use the SLD bound, since we know that the SLD translates directly to a measurement operator, and we carefully consider the attainability of the QCR bound.

\section{The model}\label{sec:model}

We consider signals with a Gaussian envelope in frequency, which achieves the minimum duration-bandwidth product $\sigma_t \sigma_\omega = 1/2$ \cite{diels2006ultrashort}.
By considering Gaussian pulses, we are able to obtain an exact analytical expression for the QFI matrix.
We expect that the same qualitative results apply to non-Gaussian pulses under fairly general assumptions.

Our theory is defined in the regime of highly attenuated signals, where each pulse contains at most one photon.
A Gaussian pulse with central frequency $\bar\omega_0$, central time $\bar t_0$, and frequency bandwidth $\sigma_0$, is described by the following single-photon wave function in the frequency representation:
\begin{align} 
 |\psi_0\rangle = \int d\omega \, \tilde \psi_0(\omega) |\omega\rangle \, ,
\end{align}
where $|\omega\rangle$ is the single photon state with frequency $\omega$, and 
\begin{align} \label{eq:psfomega}
 \tilde \psi_0(\omega) = 
 \left( \frac{1}{2 \pi \sigma_0^2} \right)^{\frac{1}{4}} 
 \exp{ \left[ -\frac{(\omega-\bar\omega_0)^2}{4\sigma_0^2} + i \omega \bar t_0 \right] }\, .
\end{align}
Alternatively, we can represent the single-photon wave function in the time domain as
\begin{align} 
 |\psi_0\rangle = \int dt \, \psi(t) |t\rangle \, ,
\end{align}
where $\ket{t}$ is the single photon state at time $t$, and 
\begin{align}\label{eq:ngbehu34iowjes}
 \psi_0(t)  & = \left( \frac{2 \sigma_0^2 }{\pi } \right)^{\frac{1}{4}} 
 \exp{ \left[ - (t-\bar t_0)^2 \sigma_0^2 - i \bar\omega_0 (t- \bar t_0) \right] } \, .
\end{align}
This signal is sent towards a target at distance $x$ that is moving with radial velocity $v$ (we choose $v$ positive when the target is moving away). 
If the photon in state $\ket{t}$ is back-scattered by the target, it is received at time
\begin{align}\label{t-delay}
 \tau(t) = t + \frac{2x}{c(1-\beta)} + \frac{2\beta(t- \bar t_0)}{1-\beta}\, ,
\end{align}
where $\beta = v/c$, and $x$ is the position of the target at time $\bar t_0$.
Therefore, the returned photon is described by the wave function
\begin{align}  \label{eq:returnunet}
 |\psi\rangle = \int dt \, \psi_0(t) | \tau(t) \rangle= \int d\tau \, \psi(\tau) | \tau \rangle \, ,
\end{align}
where
\begin{align}
 \psi(\tau) 
 & = \mathcal{N} \exp{ \left[ - \left( \frac{1-\beta}{1+\beta} (\tau-t_0) - \frac{2x}{c(1+\beta)} \right)^2 \sigma^2  \right] } \nonumber \\
 & \phantom{=}~ \times \exp{ \left[ - i \omega_0 \left( \frac{1-\beta}{1+\beta} (\tau-t_0) - \frac{2x}{c(1+\beta)} \right) \right] } \\
 & = \left( \frac{2 \sigma^2}{\pi} \right)^{\frac{1}{4}} \exp{ \left[ - \left( \tau - \bar t \right)^2 \sigma^2  - i \bar\omega \left( \tau - \bar t \right) \right] }
  \, .\label{eq:njfbher9wuoi}
\end{align}
The latter describes a Gaussian wave function with bandwidth
\begin{align}\label{bw-shift}
 \sigma = \frac{1-\beta}{1+\beta} \, \sigma_0 \, ,
\end{align}
and central time and frequency
\begin{align}
 \bar t & = \bar t_0 +  \frac{2x}{c(1-\beta)}  \, , \label{t-shift} \\ 
 \bar \omega & = \frac{1-\beta}{1+\beta} \, \bar\omega_0 \, . \label{omega-shift} 
\end{align}
The above shows that the information about the target distance $x$ is carried by central time $\bar t$, and the information about the target velocity is carried by all three parameters $\bar t$, $\bar \omega$, and $\sigma$.
The same relations hold for the case of entanglement-assisted estimation.

For entanglement-assisted sensing, we consider a model of a two-photon state as it is produced by spontaneous parametric down-conversion \cite{PhysRevLett.120.053601},
\begin{align} \label{eq:spdc}
 \ket{\Psi_0} = \int d \omega \int d \omega_i ~  
 \tilde \Psi_0(\omega,\omega_i)  \ket{\omega}\ket{\omega_i} \, , 
\end{align}
where
\begin{align} 
 \tilde \Psi_0(\omega,\omega_i) & = \tilde{\mathcal{N}}_0 \, e^{i ( \omega + \omega_i ) \bar t_0} \nonumber \\
 & \phantom{=}~ \times \exp \left[ - \frac{(\omega - \bar\omega_{0})^2}{4 (1-\kappa^2) \sigma_{0}^2} - \frac{(\omega_i-\bar\omega_{i0})^2}{4 (1-\kappa^2) \sigma_{i0} ^2} \right. \nonumber \\
 & \hspace{1.5cm} \left. - \frac{ \kappa (\omega - \bar\omega_{0}) (\omega_i-\bar\omega_{i0}) }{2(1-\kappa^2) \sigma_{0} \sigma_{i0}} \right] \, ,
\end{align}
and $\tilde{\mathcal{N}}_0$ is the the normalization factor.
This two-photon wave function describes a pair of frequency-entangled photons, with central time $\bar t_0$, central frequency $\bar \omega_{0}$, $\bar \omega_{i0}$, and bandwidth $\sigma_{0}$, $\sigma_{i0}$, for the signal and idler photons respectively.
The parameter $\kappa \in [0,1)$ quantifies the amount of entanglement between the signal and idler photon. When $\kappa=0$, the photon pair is separable, whereas in the limit when $\kappa \rightarrow 1$ the photons are perfectly entangled in frequency. Note that the state is non-physical for $\kappa =1$.

In the time domain, the two-photon wave function is
\begin{align} \label{eq:spdc-time0}
 \ket{\Psi_0} = \int d t \int d t_i ~ \Psi_0(t,t_i)  \ket{t}\ket{t_i} \, , 
\end{align}
with 
\begin{align} 
 \Psi_0(t,t_i) & = \mathcal{N}_0 \, e^{- i \bar \omega_{0} (t - \bar t_0) - i \bar\omega_{i0} (t_i - \bar t_0)} \nonumber \\
 & \phantom{=}~ \times \exp \left[ - (t - \bar t_0)^2 \sigma_{0}^2  - (t_i - \bar t_0)^2 \sigma_{i0}^2 \right. \nonumber \\
 & \hspace{1.5cm} \left. + 2 \kappa (t - \bar t_0) (t_i - \bar t_0) \sigma_{0} \sigma_{i0} \right] \, .
\end{align}
The signal photon is sent towards the target, and the idler is retained, similar to quantum illumination \cite{lloyd2008enhanced,PhysRevLett.101.253601,PhysRevLett.114.080503}. 
If the signal photon is back-scattered by the target, it will return with a time delay given by Eq.\ (\ref{t-delay}). 
By proceeding as in the single-photon case, we obtain the two-photon wave function when the returning photon is collected at the receiver:
\begin{align} \label{eq:spdc-time0}
 \ket{\Psi} = \int d t \int d t_i ~ \Psi(t,t_i)  \ket{t}\ket{t_i}
 \end{align}
where
\begin{align} 
 \Psi(t,t_i) & = \mathcal{N} e^{- i \bar\omega (t - \bar t) - i \bar\omega_{i0} (t_i - \bar t_0)} \nonumber \\
 & \phantom{=}~ \times \exp \left[ - (t - \bar t)^2 {\sigma }^2 - (t_i- \bar t_{i0})^2 \sigma_{i0}^2 \right. \nonumber \\
 & \hspace{1.5cm} \left. + 2 \kappa (t - \bar t) (t_i- \bar t_{i0}) \sigma \sigma_{i0} \right] \, ,
 \label{eq:njvhureowi}
\end{align}
and the central time $\bar t$, central frequency $\bar\omega$, and bandwidth $\sigma$ of the signal photon are as in Eqs.\ (\ref{bw-shift})-(\ref{omega-shift}).
These are the spatio-temporal properties of the back-scattered light, which we will use to extract the range and velocity of the targets.

\section{Ranging and velocity estimation}\label{Sec:1target}

In this section we present the QFI matrix for the estimation of ranging and frequency, with and without the assistance of entanglement. 
As shown in the previous Section, in our model the information about ranging $x$ and velocity $\beta$ (in natural units) of the target is carried by the central time $\bar t$, the central frequency $\bar\omega$, and the bandwidth $\sigma$ of the returned photon. 
We will first compute the QFI matrix for the estimation of the parameters $\bm{\lambda} = (\bar t_, \bar\omega,\sigma)$, and then obtain the QFI matrix for the parameters $\bm{\mu} = ( x , \beta )$ that are ultimately of interest. 
In order to find the ultimate precision of the $\bm{\mu}$-parameters, we will translate their SLDs into the SLDs for the $\bm{\lambda}$-parameters. 
We will then work with the $\bm{\lambda}$-parameters throughout the remainder of the paper.

The SLDs for the $\bm{\mu}$-parameters are related to the SLDs for the $\bm{\lambda}$-parameters as follows:
\begin{align}
    L_{\mu_j} = \sum_k \frac{\partial \lambda_k}{\partial \mu_j} L_{\lambda_k} \, ,
\end{align}
from which we obtain the QFI matrix
\begin{align}
J(\bm{\mu})_{ij} = \sum_{h,k} \frac{\partial \lambda_h}{\partial \mu_i} \frac{\partial \lambda_k}{\partial \mu_j} 
J(\bm{\lambda})_{hk} \, .
\end{align}
The partial derivatives can be evaluated to give
\begin{align}
    \frac{\partial \bar t}{\partial x} & = \frac{2}{c(1-\beta)} \, , \qquad
    \frac{\partial \bar t}{\partial \beta} = \frac{2 x }{c(1-\beta)^2} \, ,\\
    \frac{\partial \bar \omega}{\partial x} & = 0 \, , \qquad
    \frac{\partial \bar \omega}{\partial \beta} = - \frac{2 \bar \omega_0}{(1-\beta)^2} \, , \\
    \frac{\partial \sigma}{\partial x} & = 0 \, , \qquad
    \frac{\partial \sigma}{\partial \beta} = - \frac{2 \sigma}{(1-\beta)^2} \, ,
\end{align}
which yields
\begin{align}
    L_x & 
    = \frac{2}{c(1-\beta)} L_{\bar t} \, , \\
    L_\beta & 
    = \frac{2}{(1-\beta)^2} \left( \frac{x}{c} L_{\bar t} 
    - \bar\omega_0 L_{\bar \omega} 
    - \sigma L_\sigma \right) \, .
\end{align}
Note that these SLDs depend on the central time $\bar t$, the central frequency $\bar \omega$, and the bandwidth $\sigma$ of the back-scattered photon. Therefore, estimation of $x$ and $\beta$ requires that we estimate all three $\bm{\lambda}$ parameters.
In the following subsections, we will consider this estimation problem with separable and entangled photons.

\subsection{Separable photons}

We first consider the estimation of the parameters $\bm{\lambda} = (\bar t, \bar \omega,\sigma)$ using for the outgoing single-photon the wave function determined by Eq.~(\ref{eq:ngbehu34iowjes}). The back-scattered photon will have the form given in Eq.~(\ref{eq:njfbher9wuoi}).
This single-photon wave function lives in an infinite-dimensional Hilbert space. However, as we show in Appendix \ref{sec:unentangledcentral}, we can compute the QFI matrix by restricting the state to a three-dimensional Hilbert space. We define a suitable system of three basis vectors, $|e_1\rangle = |\psi\rangle$ determined by Eq.~(\ref{eq:njfbher9wuoi}), $|e_2\rangle$, and $|e_3\rangle$. In this basis, we obtain the following expression for the SLDs:
\begin{align}
L_{\bar t} &= 
\left(
\begin{array}{ccc}
 0        & 2 \sigma & 0 \\
 2 \sigma & 0        & 0 \\
 0        & 0        & 0 
\end{array}
\right) \, , \, \, 
L_{\bar\omega} = 
\left(
\begin{array}{ccc}
 0         & i/\sigma & 0 \\
 -i/\sigma & 0        & 0 \\
 0         & 0        & 0
\end{array}
\right) \, , \nn
L_\sigma & =
\left(
\begin{array}{ccc}
 0               & 0 & \sqrt{2}/\sigma \\
 0               & 0 & 0               \\
 \sqrt{2}/\sigma & 0 & 0
\end{array}
\right) \, .
\end{align}
From these SLDs, we obtain the QFI matrix
\begin{align}
J(\bm{\lambda}) = \left(
\begin{array}{cc c}
 4 \sigma^2 & 0                   & 0                  \\
 0                  & 1/ \sigma^2 & 0                  \\
 0                  & 0                   & 2/\sigma^2 
\end{array}
\right) \, .\label{eq:nebghy3iruwe}
\end{align}
As expected, the QFI of the parameter $\bar t$ is inversely proportional to that of $\bar \omega$, and it is not possible to make all diagonal elements of $J$ arbitrarily large simultaneously. The trade-off between time and frequency estimation is expressed by the relation $J(\bar t)J(\bar \omega) = 4$.

Since $\rho = |\psi\rangle \langle \psi | = | e_1 \rangle \langle e_1 |$, the necessary and sufficient condition for joint optimal estimation in Eq.\ (\ref{condition}) is $\langle \psi | [L_{\lambda_i},L_{\lambda_j}] | \psi \rangle = 0$.
However, we obtain
\begin{align}
\langle \psi | [L_{\bar t},L_{\bar \omega}] | \psi \rangle = - 4 i \, , 
\end{align}
which implies that it is not possible to jointly and optimally estimate the central time and frequency. However, as $\langle \psi | [L_{\bar t},L_{\sigma}] | \psi \rangle = \langle \psi | [L_{\bar \omega},L_{\sigma}] | \psi \rangle = 0$, it is possible to estimate jointly the bandwidth and central frequency, or the bandwidth and the central time. 

After a change of variables, we obtain the SLDs for the position and velocity of the target,
\begin{align}
L_x &= 
\frac{2}{c(1-\beta)} \left(
\begin{array}{ccc}
 0                          & 2\sigma & 0 \\
 2 \sigma & 0                          & 0 \\
 0                         & 0                           & 0 
\end{array}
\right) \, , \\
L_\beta & = 
\frac{2}{(1-\beta)^2}\left(
\begin{array}{ccc}
 0                 & 2 \sigma x/c - i \bar\omega_0/\sigma & -\sqrt{2} \\
 2 \sigma x/c + i \bar\omega_0/\sigma & 0                & 0 \\
 -\sqrt{2}                  & 0                & 0
\end{array}
\right) \, .
\end{align}
It follows that 
\begin{align}
\langle \psi | [L_x,L_\beta] | \psi \rangle = i \frac{16 \bar\omega_0}{c(1-\beta)^3} \, .
\end{align}
Thus, the SLDs for $x$ and $\beta$ inherit the incompatibility property of the central time and frequency. 
This formally shows that it is not possible to jointly estimate $x$ and $\beta$ with separable photons using the SLD measurement operators and saturating the QCR bound.

\subsection{Entangled photons}

We consider the estimation of the parameters $\bm{\lambda} = ( \bar t , \bar \omega , \sigma)$ of the two-photon wave function $\ket{\Psi}$ determined by Eq.~(\ref{eq:njvhureowi}).
As we show in Appendix \ref{sec:entangledcentral}, the QFI matrix can be computed within a four-dimensional Hilbert space using suitable basis vectors $|e_1\rangle = |\Psi\rangle$, $|e_2\rangle$, $|e_3\rangle$, $|e_4\rangle$.
In this basis, the SLDs become
\begin{align}
L_{ \bar t }&=
\sigma \sqrt{2} \left(
\begin{array}{cccc}
 0 &  \sqrt{1-\kappa} & \sqrt{1+\kappa} & 0 \\
 \sqrt{1-\kappa} & 0 & 0 & 0 \\
 \sqrt{1+\kappa} & 0 & 0 & 0 \\
 0 & 0 & 0 & 0
\end{array}
\right) \, , \\
L_{\bar \omega } &=
\frac{1}{\sigma \sqrt{2}}
\left( \begin{array}{cccc}
 0    & \frac{i}{\sqrt{1-k}}  & \frac{i}{\sqrt{1+k}} & 0 \\
 -\frac{i}{\sqrt{1-k}}   & 0 & 0   & 0 \\
 -\frac{i}{\sqrt{1+k}}  & 0 & 0   & 0 \\
  0   & 0 & 0   & 0
\end{array}
\right) \, , \\
L_{\sigma} &=
\frac{1}{\sigma} \sqrt{\frac{2- \kappa^2}{1-\kappa^2}} \left(
\begin{array}{cccc}
 0 & 0 & 0 & 1 \\
 0 & 0 & 0 & 0 \\
 0 & 0 & 0 & 0 \\
 1 & 0 & 0 & 0 
\end{array}
\right) \, .
\end{align}
For the saturability condition, we obtain
\begin{align}\label{eq:vnrhuw49eois}
\langle \Psi | [ L_{ \bar t },L_{\bar \omega } ] | \Psi \rangle = -4i \, ,
\end{align}
which is the same as for separable photons.
Therefore, the QCR bound cannot be saturated for any $\kappa \in [0,1)$. 
However, there exists a measurement, not based on the SLDs, which saturates the QCR bound in the limit where $\kappa\to 1$. This measurement was constructed by Zhuang \textit{et al.}\ \cite{PhysRevA.96.040304}, and we present this measurement using our approach in Appendix~\ref{Sec:sim_measure}.

From the SLDs we obtain the QFI matrix 
\begin{align}
 J(\bm{\lambda}) = \left(
\begin{array}{ccc}
 4\sigma^2 & 0 & 0\\
 0 & \frac{1}{\sigma^2} \frac{1}{1-\kappa^2} &0\\
 0& 0& \frac{1}{\sigma^2} \frac{2-\kappa^2}{1-\kappa^2}
\end{array}
\right) \, .
\end{align}
Note that the degree of correlation $1-\kappa^2$ appears in the denominator of the QFI for $\bar \omega$. This means that the trade-off in precision between $\bar t$ and $\bar \omega$ can be lifted by choosing $\kappa$ arbitrarily close to 1. Moreover, we can make all diagonal elements of $J$ arbitrarily large simultaneously, contrary to the case in Eq.~(\ref{eq:nebghy3iruwe}).
With the assistance of entanglement we thus obtain
\begin{align}
 J( \bar t) J(\bar \omega) = \frac{4}{1-\kappa^2} \, .
\end{align}
As previously noted in Ref.\ \cite{PhysRevA.96.040304}, this is a violation of the Arthurs-Kelly uncertainty relation \cite{arthurs1965bstj} for non-entangled photons.

\begin{figure}[t]
\includegraphics[trim = 0cm 0.5cm 0cm 0cm, clip, width=0.95\linewidth]{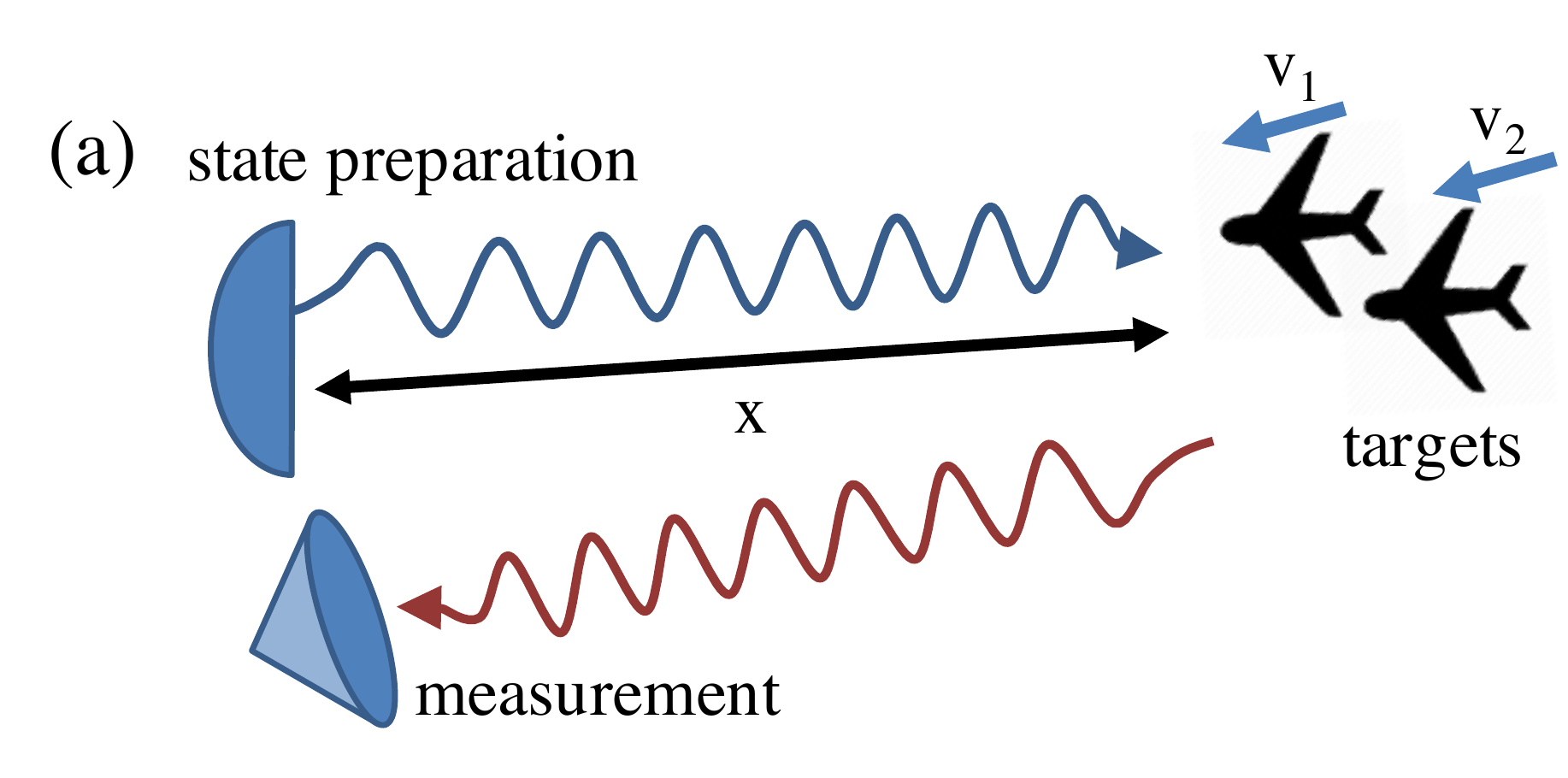} 
\includegraphics[trim = 0cm 0.5cm 0cm 0cm, clip, width=0.95\linewidth]{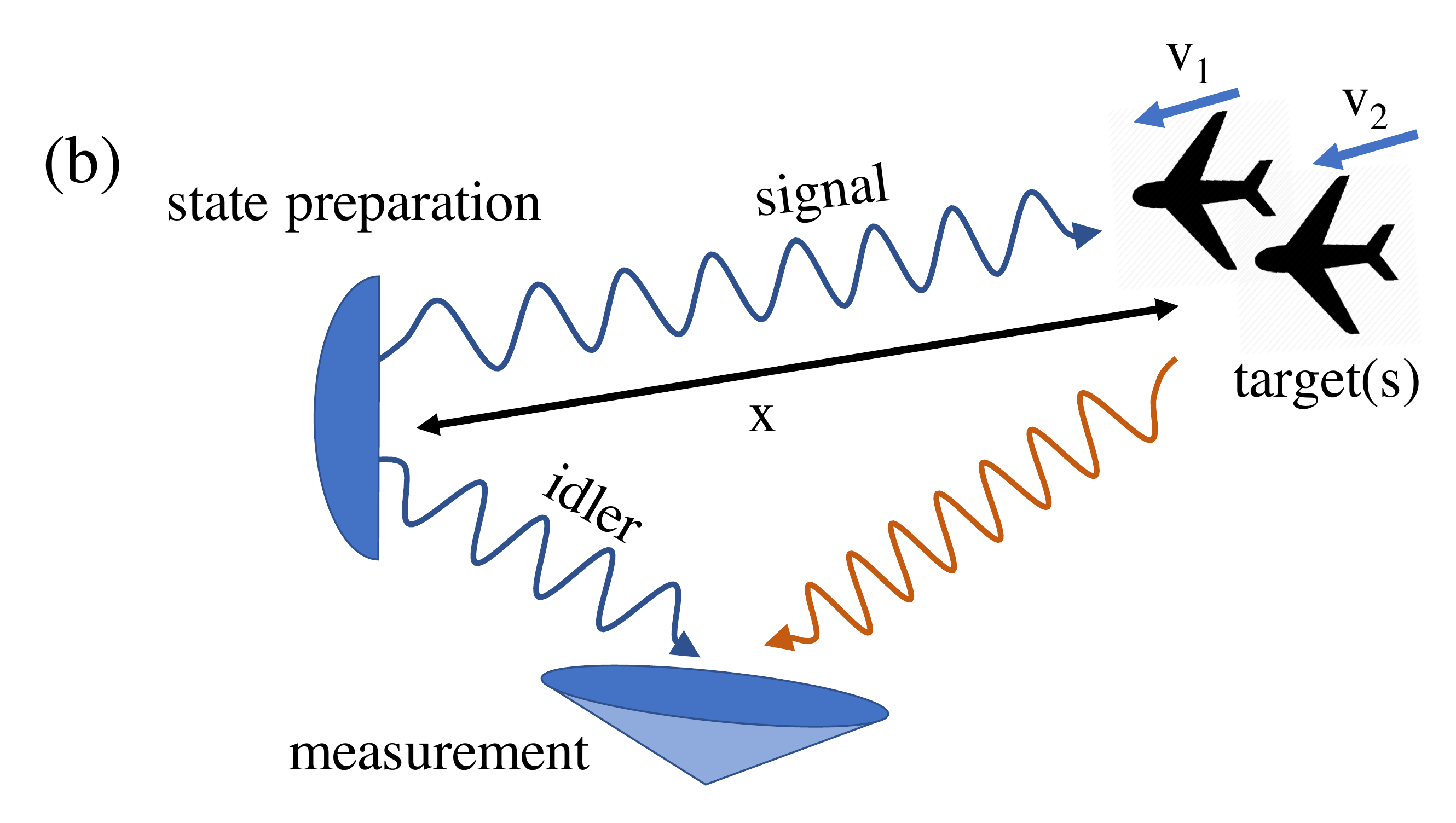}
 \caption{\label{f:scheme_2} Radar and lidar systems for measuring the velocity and range separation using (a) separable and (b) entangled states. 
}
\end{figure}

\section{Two targets}\label{Sec:2targets}

In this section we focus on the problem of estimating the relative radial position and velocity of two close targets, with either separable (Fig.\ \ref{f:scheme_2}a) or entangled (Fig.\ \ref{f:scheme_2}b) photons.

In traditional radar and lidar systems, the ability to measure the separation between two close targets deteriorates if the signals start to overlap. 
In imaging, the analogous problem, which has been dubbed \textit{Rayleigh’s curse} \cite{PhysRevX.6.031033}, arises when two objects are closer than the Rayleigh length of the optical imaging system.
In principle, if the returning signals coherently reflect off the targets, such problems can be bypassed \cite{komissarov2019partially}. 
However, this may not always be the case, especially if the pulses used are in the optical domain. 
On the other hand, it has been shown that Rayleigh's curse is an artefact of measuring only the intensity of the field, and can be avoided by using a suitable coherent detection technique \cite{PhysRevX.6.031033, PhysRevLett.124.080503}.

The same feature holds for our setting of quantum-limited lidar (and, in principle for radar too), as we will now show by computing explicitly the QFI matrix for the estimation of the relative distance and velocity of two targets.

We consider a simplified model where the information about the position and velocity of the target is only carried by the central time and central frequency, thus neglecting the bandwidth change.
This is justified by the fact that the physics is essentially determined by these two parameters only, as we have shown in detail for the case of a single target.

Assume that target $1$ has position $x_1$ and velocity $\beta_1$, and target $2$ has position $x_2$ and velocity $\beta_2$. If the photon is scattered by target $j$, it will return with central time $\bar t_j$ and central frequency $\bar \omega_j$. From Eqs.\ (\ref{t-shift})-(\ref{omega-shift}), we obtain
\begin{align}
 \bar t_j & = \bar t_{0} +  \frac{2x_j}{c(1-\beta_j)}  \, , \\ 
 \bar \omega_j & = \frac{1-\beta_j}{1+\beta_j} \, \bar\omega_{0} \, .
\end{align}
Therefore
\begin{align}
 \Delta t : = \bar t_1 - \bar t_2 & = \frac{2x_1}{c(1-\beta_1)} - \frac{2x_2}{c(1-\beta_2)}  \\
 & \simeq \frac{2(x_1-x_2)}{c}  \, , \\
\Delta \omega : = \bar \omega_1 - \bar \omega_2 & = \frac{1-\beta_1}{1+\beta_1} \, \bar\omega_{0}
 - \frac{1-\beta_2}{1+\beta_2} \, \bar\omega_{0} \\
& \simeq - 2(\beta_1 - \beta_2) \, \bar\omega_{0} \, ,
\end{align}
where the approximations hold in the non-relativistic regime $\beta_1,\beta_2 \ll 1$.
Putting $\Delta x := x_1 - x_2$ and $\Delta \beta := \beta_1 - \beta_2$, we obtain 
\begin{align}
    \frac{\partial \Delta t}{\partial \Delta x} = \frac{2}{c} \, , \qquad
    \frac{\partial \Delta \omega}{\partial \Delta \beta} = - 2 \bar \omega_0 \, .
\end{align}
This allows us to write the SLDs for the parameters $\Delta x$, $\Delta \beta$ in terms of the SLDs for $\Delta t$ and $\Delta\omega$:
\begin{align}
    L_{\Delta x} = \frac{2}{c} L_{\Delta t} \, , \qquad L_{\Delta \beta} = - 2 \bar \omega_0 L_{\Delta \omega} \, .
\end{align}
Below, we first compute the QFI for separable photons, and then consider the use of entangled photon pairs.

\subsection{Separable photons}

\begin{figure}[t!]
\includegraphics[trim = 0cm 0.5cm 0cm 0cm, clip, width=0.8\linewidth]{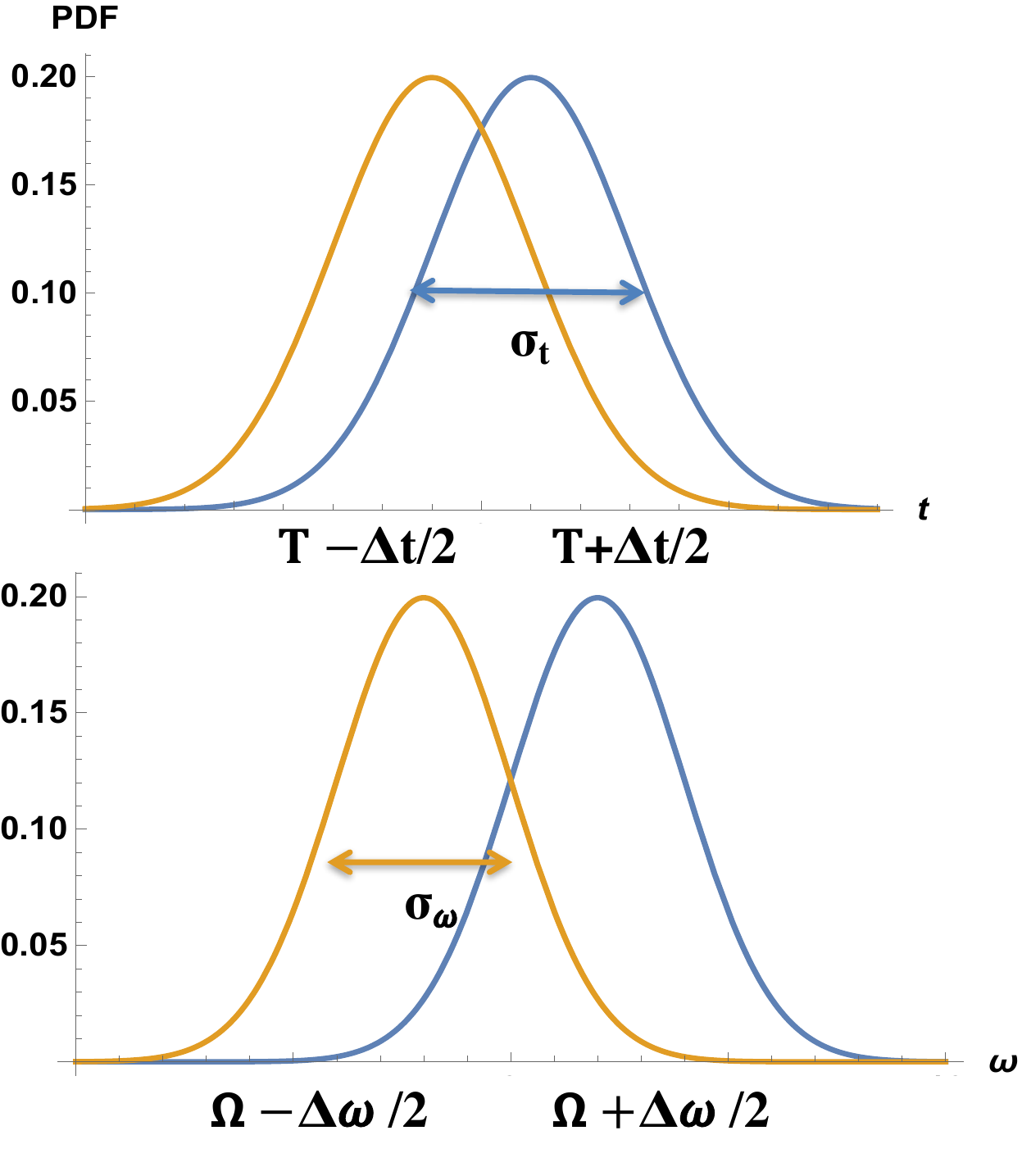} 
 \caption{Time and frequency profiles of of arrival of a single photon signal scattering off two objects within the vicinity of each other, where their separation is within the bandwidth of the pulse.} \label{f:time_profile}
\end{figure}

In this section we consider an outgoing single-photon wave function determined by Eq.~(\ref{eq:ngbehu34iowjes}). The back-scattered photon will have the form given in Eq.~(\ref{eq:njfbher9wuoi}). 
If the photon returns to the detector, it means it has been back-scattered by either target $1$ or target $2$. As the scattering events are assumed to be incoherent, the returned photon is described by the mixed state
\begin{align} \label{eq:rho_une}
\rho = \frac{1}{2} \ket{\psi_1}\bra{\psi_1} + \frac{1}{2} \ket{\psi_2}\bra{\psi_2} \, ,
\end{align}
where we assume the reflectivities of the two objects are approximately equal.
We expect that our results hold also for unequal reflectivities \cite{PhysRevLett.124.080503}.
We thus use
$|\psi_j\rangle = \int dt\, \psi_j(t) |t\rangle$, for $j=1,2$, with
\begin{align}
\psi_j(t)  = \left( \frac{2 \sigma^2 }{\pi } \right)^{\frac{1}{4}} 
\exp{ \left[ - (t - \bar t_{j})^2 \sigma^2 - i \bar \omega_{j} (t- \bar t_{j}) \right] }
\, .
\end{align}
We define the centroids in time ($T$) and frequency ($\Omega$) as (see Fig.\ \ref{f:time_profile})
\begin{align}
 T = \frac{ \bar t_{1} + \bar t_{2}}{2} \, , \qquad
 \Omega = \frac{\bar\omega_{1} + \bar\omega_{2}}{2} \, ,
\end{align}
The goal of this Section is to compute the QFI matrix for the estimation of the parameters $\Delta t$ and $\Delta \omega$.

\begin{figure}[t]
\includegraphics[trim = 0cm 0cm 0.0cm 0.0cm, clip, width=0.95\linewidth]{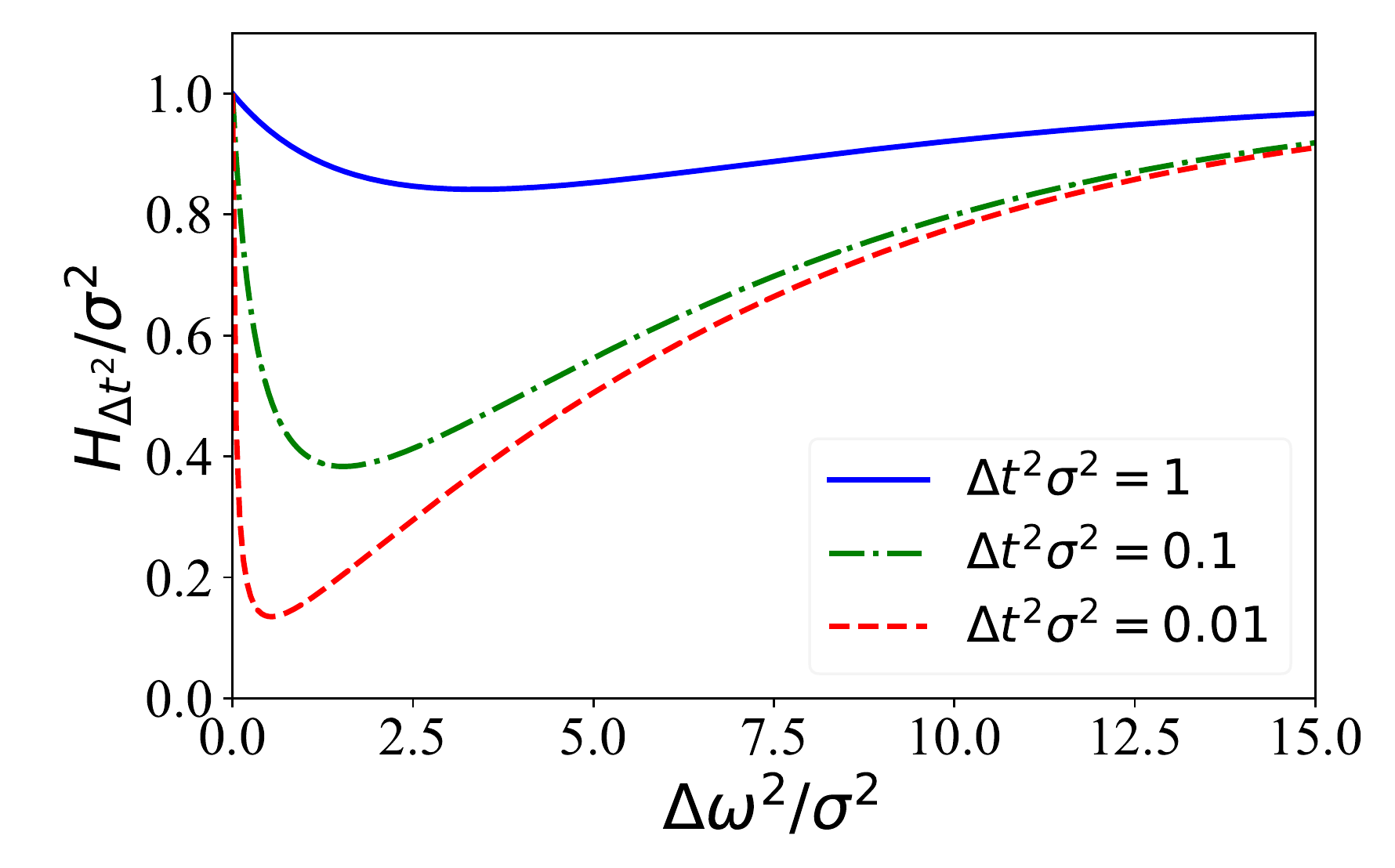} 
 \caption{\label{f:hdt} QFI matrix component $H_{\Delta t^2}$ in Eq.~\eqref{eq:qfi_unentangled} as a function of 
 $\Delta \omega^2/\sigma^2$, showing $\Delta t^2 \sigma^2 = 0.01$ (red dashed line), 
 $\Delta t^2 \sigma^2 =0.1$ (green dotted dashed line), and $\Delta t^2 \sigma^2 =1$ (blue solid line).  }
\end{figure}

Following Ref.\ \cite{PhysRevX.6.031033}, we obtain an exact expression for the QFI matrix for the estimation of $\Delta t$ and $\Delta \omega$
\begin{align} \label{eq:qfi_unentangled}
H(\Delta t,\Delta\omega) & =
\left(
\begin{array}{cc}
 \sigma^2 - \alpha^{-1} \Delta \omega^2 &  \alpha^{-1} \Delta t \Delta \omega \\
 \alpha^{-1} \Delta t \Delta \omega & \frac{1}{4 \sigma^2} - \alpha^{-1} \Delta t^2 
\end{array}
\right) \, , \\
\alpha & = 4 
\left(e^{\Delta t^2 \sigma^2 + \frac{\Delta \omega ^2}{4 \sigma^2}}-1 \right) \, ,
\end{align}
see Appendix \ref{app:bvhreu9j}. 
The expectation value of the commutator of the SLDs becomes 
\begin{align}
    \text{Tr}( \rho [L_{\Delta t},L_{\Delta \omega}] ) & =\frac{4i}{\alpha} \left( \Delta t^2 \sigma^2+\frac{\Delta \omega ^2}{4 \sigma^2}\right) -i \\
    & \simeq - \frac{i}{2} \left( \Delta t^2 \sigma^2+\frac{\Delta \omega ^2}{4 \sigma^2}\right) \, ,
\end{align}
where the approximation holds, for small values of $\epsilon = \Delta t^2 \sigma^2+\frac{\Delta \omega ^2}{4 \sigma^2}$, up to correction of order $\epsilon^2$.
For small values of $\Delta t \sigma$ and $\Delta\omega/\sigma$ this quantity approaches zero, and therefore the achievable estimation precision approaches the QCR bound. 
Note that in this limit the QFI matrix becomes diagonal.
This is in contrast to the single target ranging problem from the previous section, where the expectation value of the commutator of the SLDs was a constant $-4i$, see Eq.~(\ref{eq:vnrhuw49eois}).
Some values of $H_{\Delta t^2}$ as a function of $\Delta\omega^2$ are shown in Fig.~\ref{f:hdt}. 

\subsection{Entangled photons}

Next, we consider again the case where the probe photon is entangled with an idler photon, for example when both are created in spontaneous parametric down-conversion.
If the back-scattering is again incoherent, the two-photon state becomes
\begin{align} \label{eq:rho}
\rho = \frac{1}{2} \ket{\Psi_1}\bra{\Psi_1} + \frac{1}{2} \ket{\Psi_2}\bra{\Psi_2} \, ,
\end{align}
where $\ket{\Psi_j} = \int dt\, dt_i \Psi_j(t,t_i) |t\rangle |t_i\rangle$, $i$ denotes the idler photon, $j = 1,2$, and 
\begin{align} 
\Psi_j(t,t_i) & = \mathcal{N} e^{- i \bar \omega_{j} (t - \bar t_{j}) - i \bar \omega_{i} (t_i - \bar t_i)} \nonumber \\
& \phantom{=}~ \times \exp \left[
- (t - \bar t_{j})^2 \sigma^2 
- (t_i - \bar t_i)^2 \sigma_{i}^2 
\right. \nonumber \\
& \hspace{1.5cm} \left. + 2 \kappa (t - \bar t_{j}) (t_i - \bar t_i) \sigma \sigma_{i} \right] \, .
\end{align}
This means that, although the entangled state lives in a larger Hilbert space, the formal approach used for separable photons can still be applied.
As detailed in Appendix \ref{sec:entangledsep}, we obtain the following expression for the QFI matrix:
\begin{align}
H_\text{ent} & =
\left(
\begin{array}{cc}
 \sigma^2 - \beta^{-1} \Delta \omega^2 & \beta^{-1} \Delta \omega \Delta t \\
 \beta^{-1} \Delta \omega \Delta t & \frac{1}{4 \sigma^2(1- \kappa^2 )} - \beta^{-1} \Delta t^2 
\end{array}
\right) \\
\beta & = 4 \left(e^{ \Delta t^2 \sigma^2+\frac{\Delta \omega^2}{4 \left(1-\kappa^2\right) \sigma^2} }-1\right) \, .
\end{align}
Here we see a similar quantitative change in the QFI matrix compared to single target detection: the term $1-\kappa^2$ appears as a multiplicative factor in front of the bandwidth $\sigma_s$, which effectively reduces the frequency uncertainty of the probe photon.
Provided $\kappa \neq 0$, the QFI matrix's components are always larger than the unentangled case. The expectation value of the commutator of the SLDs becomes
\begin{align}
    \text{Tr}( \rho [L_{\Delta t},L_{\Delta \omega}] ) & =\frac{4i}{\beta} \left( \Delta t^2 \sigma^2+\frac{\Delta \omega^2}{4 \left(1-\kappa^2\right) \sigma^2} \right) -i \\
    & \simeq - \frac{i}{2} \left( \Delta t^2 \sigma^2+\frac{\Delta \omega^2}{4 \left(1-\kappa^2\right) \sigma^2} \right) \, ,
\end{align}
which approaches zero for small values of $\Delta t\, \sigma$ and $\Delta\omega/\sigma$.

In conclusions, we found that if the targets are sufficiently close in both position and velocity, the SLD measurements become compatible up to a small eigenbasis mismatch. 
In this limit the two parameters become jointly measurable and the QCR bound can be saturated.

\section{Optimal time-difference estimation with linear optics} \label{sec:measurement}

The QFI provides us with an upper bound to the ultimate precision, but does not always provide the optimal physical measurement. 
Consider a quantum state $\rho(\lambda)$, which carries information about the parameter $\lambda$.
When a given measurement $M$ is applied to $\rho(\lambda)$, it yields outcomes $\{m_y \}_{y=1,\dots,Y}$ with probabilities $\{p_y\}_{y=1,\dots,Y}$.
The classical Fisher information (CFI) associated with this measurement is \cite{frieden2004science}
\begin{align} \label{eq:fi}
I(\lambda) = \sum_y p_y \left( \frac{\partial \log{p_y} }{\partial \lambda}\right)^2  \, . 
\end{align}
The (classical) Cram\'er-Rao bound expresses the relation between the CFI and the variance of any unbiased estimator $\hat \lambda$,
$\Delta \hat\lambda \geq \frac{1}{N} I(\lambda)^{-1}$.
An optimal measurement is such that the CFI is equal to the QFI.

Here we consider the two-target problem, and provide an optimal measurement for the estimation of $\Delta t$ when $\Delta \omega=0$, i.e.\ the two targets are moving at the same velocity. In particular, we focus on the case of separable photons, described by the state in Eq.~\eqref{eq:rho_une}.
In this setup, an optimal measurement was presented in Ref.\ \cite{PhysRevLett.121.090501} using a quantum pulse gate. Unlike the quantum pulse gate which is based on up-conversion, here we propose an approach that requires no optical non-linearity.

\begin{figure}
 \includegraphics[trim = 0.5cm 0cm 0.3cm 0cm, clip, width=1.0\linewidth]{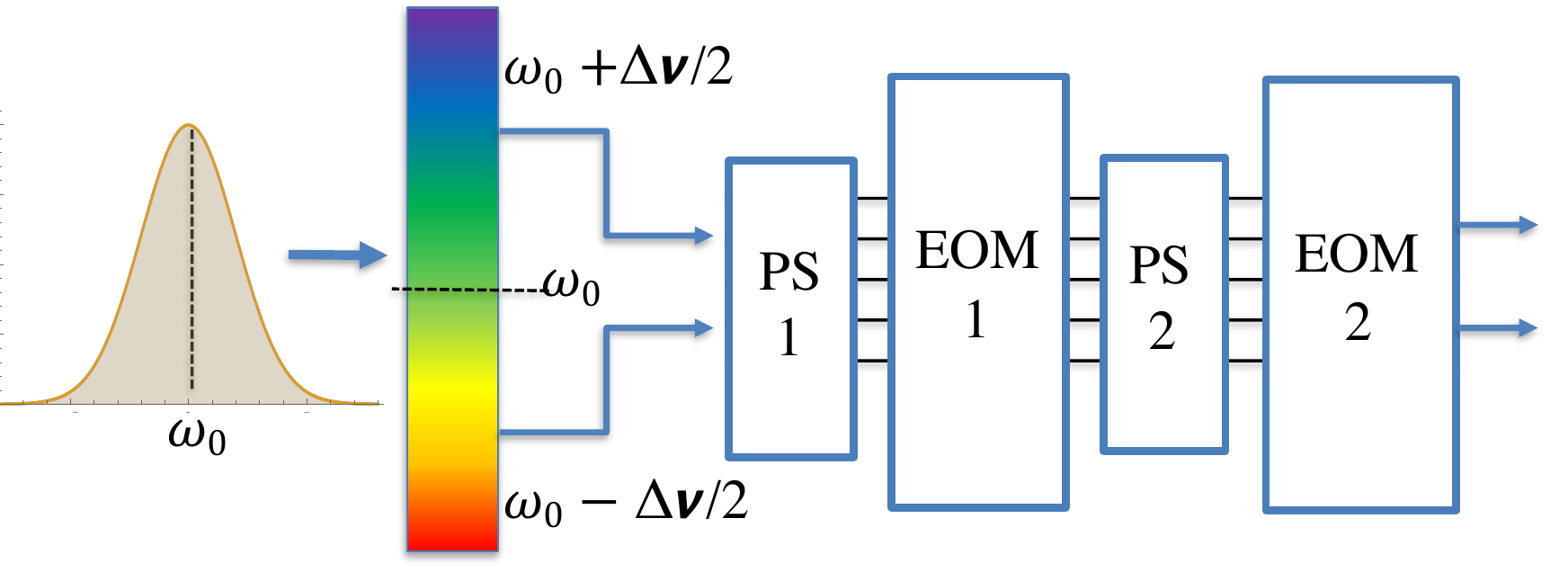} 
 \caption{
 A linear measurement which achieves the QFI for the parameter $\Delta t$, for the special case where $\Delta \omega =0$. It involves separating the signal into different frequency modes, then selecting the frequencies equidistant from either side of the central frequency, followed by a frequency Hadamard gate.
 The schematic for the Hadamard gate consists of two phase-shifters (PS) and two electro-optic modulators (EOM). Photon counting is then performed at the output.
 }\label{f:optical_hadamard} 
\end{figure}

A schematic for an optimal linear measurement is depicted in Fig.~\ref{f:optical_hadamard}.
The measurement consists of first sending the signal through a diffraction grating, which separates the frequencies within the pulse. Then, one takes the frequencies on either side of the centroid that are equidistant, and interfere them through a frequency Hadamard gate. Finally, photon counting is performed at the output.

We now derive the classical Fisher information associated with this particular measurement. Upon the signal's return from target $1$ or $2$, the annihilation operators can be written as
\begin{align}
\hat a_1(t) &= \int  d\omega ~\hat a(\omega) ~e^{-i \omega t_1 + i\phi_{k0}} \, , \\
\hat a_2(t) &=  \int  d\omega ~\hat a(\omega) ~ e^{-i \omega  t_2 + i\phi_{k0}} \, .
\end{align}
\noindent Here we choose $\phi_{k0}=0$ without loss of generality.
Now, we select the frequencies at either side of the central frequency $\bar \omega$ separated by $\Delta \nu$: 
\mbox{$\nu_1 = \bar \omega + \Delta \nu/2$} and 
\mbox{$\nu_2 =\bar \omega - \Delta \nu/2$}. Postselected on these two frequencies, the density matrix of the state upon return can be written as
\begin{align}
\rho &= \frac{1}{2}(\ket{f_1}\bra{f_1} + \ket{f_2}\bra{f_2}) \, , \\
\ket{f_1} &= \frac{1}{\sqrt2}(\ket{\nu_1} + e^{i (\nu_2-\nu_1) t_1}\ket{\nu_2}) \, , \\
\ket{f_2} &= \frac{1}{\sqrt2}(\ket{\nu_1} + e^{i (\nu_2-\nu_1) t_2}\ket{\nu_2}) \, .
\end{align}

The experimenter implements a frequency Hadamard gate on the states $\ket{\nu_1}$ and $\ket{\nu_2}$, which is achievable using two phase shifters and electro-optical modulators, i.e., without employing non-linear optics.
Such a Hadamard gate has been experimentally demonstrated in Ref.~\cite{lukens2017frequency} with unit success probability and fidelity. The Hadamard gate is given by
\begin{align}
    H = \frac{1}{\sqrt{2}}
\left(
\begin{array}{cc}
 1 & 1 \\
 1 & -1 \\
\end{array}
\right).
\end{align}
The state then becomes
\begin{align}
    \rho' = \left(
\begin{array}{cc}
 p_1 & b \\
 b^* & p_2 \\
\end{array}
\right),
\end{align}
\noindent where $b = \frac{1}{2} \sin (t_2 (\nu_1-\nu_2)) - \frac{i}{2} \sin (t_1 (\nu_1-\nu_2)$.
The diagonal terms give the probabilities of the two measurement outcomes at the output of the Hadamard gate,
\begin{align} \label{eq:outcomesingle}
p_1 &=\frac{1}{4} \left( 2+\cos [(\nu_2-\nu_1) t_1]+\cos [(\nu_2-\nu_1)  t_2] \right) \, , \\
p_2 &=\frac{1}{4} \left( 2-\cos [(\nu_2-\nu_1)   t_1]-\cos [(\nu_2-\nu_1)  t_2] \right) \, .
\end{align}
\noindent
The Fisher information of the parameter $\Delta t = t_1 - t_2$ for Eq.~\eqref{eq:outcomesingle} is
\begin{align} \label{eq:FIsigle}
I(\Delta t)
           =  (\nu_1-\nu_2)^2/4 \, .
\end{align}
%
To calculate the overall Fisher information, we need to average Eq.~\eqref{eq:FIsigle} over the frequency distribution.
The returning state has a probability density distribution (PDF)
\begin{align}
p_\Omega(\omega - \bar \omega) &= \left( \frac{1}{2 \pi \sigma^2} \right)^{\frac{1}{2}} 
\exp{ \left[ -\frac{(\omega-\bar\omega)^2}{2\sigma^2} \right] } \, ,
\end{align}
\noindent and the PDF of $|\nu_1-\nu_2|$ is equal to $p_\Omega((\nu_1-\nu_2)/2)$.
Using Eq.~\eqref{eq:fi}, the expression for the CFI is 
\begin{align}\label{eq:averaged}
I(\Delta t) &=\int_{0}^{\infty} p_\Omega (( \nu_1-\nu_2)/2)  \frac{| \nu_1-\nu_2|^2}{4}  
  =  \sigma^2 \, .
\end{align}
\noindent Eq.~\eqref{eq:averaged} is equal to the corresponding element of the QFI in Eq.~(\ref{eq:qfi_unentangled}), and thus the measurement is optimal.

Intuitively, the frequency bandwidth of the state plays the same role as the numerical aperture (i.e., the size of the lens) in classical optical imaging.
For optical imaging in the paraxial regime, the source distribution is Fourier-transformed into its spatial frequency components. The larger the numerical aperture, the more information one can collect on the source's spatial distribution.  Likewise, if the goal is to collect timing information, the larger the frequency bandwidth, the more accurately one can locate the pulses in the time domain. In both cases, using a phase-sensitive measurement instead of the intensity profile can avoid Rayleigh's curse.

\section{Conclusions}

In this paper, we have assessed the ultimate precision of lidar or radar systems using the theoretical toolbox of multiparameter quantum metrology.
We have considered both the case of a single target as well as a pair of close targets. In the latter case, we have focused on the problem of estimating the relative position and velocity.

%

Our theory shows that the trade-off between the estimation of time and frequency can be weakened when the signal photon is entangled and jointly measured with an idler photon. In other words, the bandwidth-duration product is no longer lower bounded by 1/2.
The more the photon pair is entangled, the more the trade-off is weakened, and it can be completely removed in the limit of infinite entanglement. In that case the QCR bound is attainable.
Our results are consistent with what was previously presented by Zhuang, Zhang, and Shapiro \cite{PhysRevA.96.040304}, and elucidate the subtle issues surrounding the estimation of time and frequency.

%

 
%
%

For the estimation of the relative distance and velocity of two targets, one expects that the precision deteriorates when the two targets are close enough, such that there is a substantial overlap between the two back-scattered signals.
This is the lidar analogy of the so-called \textit{Rayleigh curse}, which is observed in classical optical imaging based on direct photo-detection \cite{PhysRevX.6.031033}. 
In previous work, Silberhorn and collaborators have shown that a coherent detection technique allows us to lift  Rayleigh’s curse, and measure the difference in time of arrival with constant precision, independent of the signal overlap \cite{PhysRevLett.121.090501,ansari2020achieving}. The same holds for the estimation of time or frequency difference.

Here we considered the joint estimation of both the difference in time of arrival (i.e., the relative position of two targets) and in frequency (i.e., the relative velocity of two targets).
In analogy with the case of a single target, we found that there exists a trade-off between time and frequency difference estimation, and that this trade-off can be weakened if entangled photon pairs are employed. 
However, in contrast to the single target case and classical intuition, we have shown that these parameters can be simultaneously estimated, even without entanglement, in the regime where the two signals have a substantial overlap, that is, when the frequency difference is much smaller than the bandwidth and the relative time of arrival is much smaller than the signal duration.
Our results will be important for the realistic implementation of super-resolution lidar systems with finite entanglement in the probe beam.


\begin{acknowledgements}
ZH thanks Lorenzo Maccone and Ilaria Gianani for insightful discussions. This work was supported by the EPSRC Quantum Communications Hub, Grant No.EP/M013472/1. 
\end{acknowledgements}

\widetext

\appendix

\section{Ranging and velocity estimation with separable photons}\label{sec:unentangledcentral}

For the un-entangled case, we consider the state $\ket{\psi} = \int_{-\infty}^\infty dt \, \psi(t) \ket{t}$, with
\begin{align}\label{eq:unent}
\psi(t) = \left(\frac{2\sigma^2}{\pi}\right)^{1/4}
e^{-(t - \bar t)^2 \sigma^2}
 e^{-i \bar\omega (t - \bar t)}  \, .
\end{align}

We define the vectors $\ket{\phi_k} = \int_{-\infty}^\infty dt \, \phi_k(t) \ket{t}$, for $k=1,2,3,4$, where
\begin{align}
\phi_1(t) & : = \psi(t) \, , \\
\phi_2(t) & : = \partial_{\bar t} \psi(t) 
=  \left[ 2 (t - \bar t) \sigma^2  + i \bar \omega \right] \psi(t) \, , \\
\phi_3(t) & = \partial_{\bar\omega} \psi(t) 
= - i (t - \bar t) \psi(t) \, , \\
\phi_4(t) & = \partial_\sigma \psi(t) 
= \left[ \frac{1}{2 \sigma } - 2 (t - \bar t)^2 \sigma \right] \psi(t) \, .
\end{align}
The local dynamics of the state $\ket{\psi}$, for small variations of the parameters $\bar t$, $\bar\omega$, $\sigma$ is confined within the Hilbert space generated by these four vectors. 
It is easy to check that the above vectors span a three-dimensional Hilbert space. An orthonormal basis for this space is $\ket{e_j} = \int_{-\infty}^\infty dt \, e_j(t) \ket{t}$, for $j=1,2,3$, where
\begin{align}
e_1(t) & = \psi(t) \, , \\
e_2(t) & = 2 \sigma (t- \bar t) \psi(t) \, , \\
e_3(t) & = \frac{ 1 - 4 (t- \bar t)^2 \sigma^2 }{\sqrt{2}} \, \psi(t) \, .
\end{align}

In this basis, we obtain the following expression for the SLDs:
\begin{align}
L_{\bar t} &= 
2 \sigma \left(
\begin{array}{ccc}
 0 & 1 & 0 \\
 1 & 0 & 0 \\
 0 & 0 & 0 
\end{array}
\right) \, , \\
L_{\bar \omega} & = 
\frac{1}{\sigma} \left(
\begin{array}{ccc}
 0  & i & 0 \\
 -i & 0 & 0 \\
 0  & 0 & 0
\end{array}
\right) \, , \\
L_\sigma & =
\frac{\sqrt{2}}{\sigma} \left(
\begin{array}{ccc}
 0 & 0 & 1 \\
 0 & 0 & 0 \\
 1 & 0 & 0
\end{array}
\right) \, .
\end{align}

Therefore the QFI matrix is
\begin{align}
H(\bar t, \bar \omega, \sigma) = \left(
\begin{array}{cc c}
 4 \sigma^2 & 0          & 0         \\
 0          & 1/\sigma^2 & 0         \\
 0          & 0          & 2/\sigma^2
\end{array}
\right) \, .
\end{align}

We can then obtain the SLDs for the estimation of the position $x$ and velocity $\beta$ of a moving target. We obtain:
\begin{align}
L_x &= 
\frac{2}{c(1-\beta)} \left(
\begin{array}{ccc}
 0                          & 2\sigma & 0 \\
 2 \sigma & 0                          & 0 \\
 0                         & 0                           & 0 
\end{array}
\right) \, , \\
L_\beta & = 
\frac{2}{(1-\beta)^2}\left(
\begin{array}{ccc}
 0                 & 2 \sigma x/c + i \bar\omega_0/\sigma & \sqrt{2} \\
 2 \sigma x/c - i \bar\omega_0/\sigma & 0                & 0 \\
 \sqrt{2}                  & 0                & 0
\end{array}
\right) \, ,
\end{align}
and 
\begin{align}
H(x,\beta) = \frac{4}{(1-\beta)^2} \left(
\begin{array}{cc}
 \frac{4 \sigma^2}{c^2} & \frac{4 x \sigma^2}{c^2 (1-\beta)}          \\
 \frac{4 x \sigma^2}{c^2 (1-\beta)}          &  \frac{4 x^2 \sigma^4 + c^2 ( 2 \sigma^2 + \bar\omega_0^2 ) }{c^2 \sigma^2 (1-\beta)^2} 
\end{array}
\right) \, .
\end{align}

\section{Ranging and velocity estimation using entangled photons}\label{sec:entangledcentral}

In the time domain, the two-photon wave function reads. 
\begin{align} \label{eq:spdc-time}
\ket{\Psi} = 
 \int dt \int dt_i ~ 
 \Phi(t,t_i)  \ket{t}\ket{t_i} \, , 
\end{align}
with 
\begin{align} 
\Psi(t,t_i) = e^{- i \bar\omega (t - \bar t) - i \bar\omega_{i} (t_i - \bar t_{i})} (1-\kappa ^2)^{1/4} \sqrt{\frac{2 \sigma \sigma_i}{\pi }}  
\exp & \left[
- (t- \bar t)^2 \sigma^2 
- (t_i- \bar t_{i})^2 \sigma_i^2 
\right.
\left. + 2\kappa (t- \bar t ) (t_i - \bar t_{i}) \sigma \sigma_i \right] \, .
\end{align}

Consider the following vectors:
\begin{align}
\Phi_1 (t,t_i) & : = \Psi (t,t_i) \, , \\
\Phi_2 (t,t_i) & : = \partial_{\bar t} \Psi (t,t_i) 
= \left[ i \bar\omega + 2 (t - \bar t ) \sigma^2 - 2 \kappa (t_i - \bar t_{i}) \sigma \sigma_i \right] \Psi (t,t_i) \, , \\
\Phi_3 (t,t_i) & : = \partial_{\bar\omega} \Psi (t,t_i) 
= - i (t- \bar t) \Psi (t,t_i) \, , \\
\Phi_4 (t,t_i) & : = \partial_{\sigma} \Psi (t,t_i) 
= \left[ \frac{1}{2 \sigma} - 2 (t - \bar t)^2 \sigma + 2 \kappa (t - \bar t) (t_i - \bar t_{i}) \sigma_i \right] \Psi (t,t_i) \, .
\end{align}

These vectors generate a four-dimensional Hilbert space.
An orthonormal basis for this space is $\ket{e_j} = \int_{-\infty}^\infty dt \int_{-\infty}^\infty dt_i \, e_j(t,t_i) \ket{t} \ket{t_i}$, for $j=1,2,3,4$, where
\begin{align}
e_1(t,t_i) & = \Psi(t,t_i) \, , \\
e_2(t,t_i) & = \sqrt{2(1-\kappa)} \left[ \sigma  (t - \bar t) + \sigma_i  (t_i - \bar t_{i}) \right] \Psi(t,t_i) \, , \\
e_3(t,t_i) & = \sqrt{2(1+\kappa)} \left[ \sigma (t - \bar t) - \sigma_i  (t_i - \bar t_{i}) \right]  \Psi(t,t_i) \, , \\
e_4(t,t_i) & = 2 \sigma \sqrt{\frac{ 1-\kappa^2 }{2- \kappa^2} } \left[ \frac{1}{2\sigma} - 2 (t - \bar t )^2 \sigma + 2 \kappa (t - \bar t) (t_i - \bar t_{i}) \sigma_i \right] \Psi (t,t_i) \, .
\end{align}

We then obtain the following expressions for the SLDs:
\begin{align}
L_{ \bar t} &=
\sigma \sqrt{2} \left(
\begin{array}{cccc}
 0 &  \sqrt{1-\kappa} & \sqrt{1+\kappa} & 0 \\
 \sqrt{1-\kappa} & 0 & 0 & 0 \\
 \sqrt{1+\kappa} & 0 & 0 & 0 \\
 0 & 0 & 0 & 0
\end{array}
\right) \, , \\
L_{\bar\omega} &=
\frac{1}{\sigma \sqrt{2}}
\left( \begin{array}{cccc}
 0    & \frac{i}{\sqrt{1-k}}  & \frac{i}{\sqrt{1+k}} & 0 \\
 -\frac{i}{\sqrt{1-k}}   & 0 & 0   & 0 \\
 -\frac{i}{\sqrt{1+k}}  & 0 & 0   & 0 \\
  0   & 0 & 0   & 0
\end{array}
\right) \, , \\
L_{\sigma} &=
\frac{1}{\sigma} \sqrt{\frac{2- \kappa^2}{1-\kappa^2}} \left(
\begin{array}{cccc}
 0 & 0 & 0 & 1 \\
 0 & 0 & 0 & 0 \\
 0 & 0 & 0 & 0 \\
 1 & 0 & 0 & 0 
\end{array}
\right) \, .
\end{align}

They in turn yield the QFI matrix:
\begin{align}
H( \bar t , \bar\omega, \sigma ) =
\left(
\begin{array}{ccc}
 4 \sigma^2  & 0 & 0\\
 0 & \frac{1}{\sigma^2} \frac{1}{1-\kappa^2} & 0\\
 0 & 0 & \frac{1}{\sigma^2} \frac{2-\kappa ^2}{ 1-\kappa ^2 }
\end{array}
\right) \, . 
\end{align}

\section{Relative range and velocity estimation with separable photons}\label{app:bvhreu9j}

For the un-entangled case, we consider the state 

\begin{align}
\rho =\frac{1}{2} \left( \ket{\psi_1}\bra{\psi_1} + \ket{\psi_2}\bra{\psi_2} \right) \, ,
\end{align}
where
\begin{align}
\ket{\psi_1} = \int_{-\infty}^\infty dt \, \psi_1(t) \ket{t} \ , \qquad
\ket{\psi_2} = \int_{-\infty}^\infty dt \, \psi_2(t) \ket{t} \, ,
\end{align}
and
\begin{align}\label{eq:unent}
\psi_1(t) = \left(\frac{2\sigma^2}{\pi}\right)^{1/4}
e^{-(t - \bar t_1)^2 \sigma^2}
 e^{-i \bar\omega_1 (t - \bar t_1)}  \, , \nn
 \psi_2(t) = \left(\frac{2\sigma^2}{\pi}\right)^{1/4}
e^{-(t - \bar t_2)^2 \sigma^2}
 e^{-i \bar\omega_2 (t - \bar t_2)}  \, .
\end{align}

We write the time and frequency in terms of their centroids and separations,
\begin{align}
\bar t_1 &= \Sigma t +\Delta t/2 \, , \qquad \bar\omega_1 = \Sigma \omega + \Delta \omega/2 \\
\bar t_2 &= \Sigma t -\Delta t/2 \, , \qquad \bar\omega_2 = \Sigma \omega - \Delta \omega/2 \, .
\end{align}

An orthonormal basis for this space is 
\begin{align} 
\ket{e_1} &= \frac{1}{\sqrt{2(1+|\delta|)}}(\ket{\psi_1} + e^{i\Delta t \Sigma \omega} \ket{\psi_2} ) \, , \\
\ket{e_2} &= \frac{1}{\sqrt{2(1-|\delta|)}}(\ket{\psi_1} - e^{i\Delta t \Sigma \omega} \ket{\psi_2} ) \, , \\
\ket{e_3} &= \frac{1}{\sqrt{c_3}}( \ket{\partial_{\Delta t} e_1} - 
\braket{e_1|\partial_{\Delta t} e_1}\ket{e_1} ) \, , \\
\ket{e_4} &= \frac{1}{\sqrt{c_4}}( \ket{\partial_{\Delta t} e_2} - 
\braket{e_1|\partial_{\Delta t} e_2}\ket{e_2} ) \, , 
\end{align}
where
\begin{align} 
\delta 
= \braket{\psi_1|\psi_2} 
= e^{-\frac{\Delta t^2 \sigma^2}{2} -\frac{\Delta \omega^2}{8 \sigma^2} - i \Delta t \Sigma \omega}
\, ,
\end{align}
and $c_3$ and $c_4$ are normalisation factors.
We can diagonalise the state as
\begin{align}
\rho = & ~p_1  \ket{e_1} \bra{e_1} + p_2 \ket{e_2}\bra{e_2} \, , \\
p_1 =  &  \frac{1}{2}(1+|\delta|) \, , \qquad p_2 = \frac{1}{2}(1-|\delta|) \, .
\end{align}

\noindent 
The SLD's are
\begin{align} L_{\Delta t} & =
\left(
\begin{array}{cccc}
\partial_{\Delta t} p_1/p_1 & 0 & 2\braket{\partial_{\Delta t} e_1|e_{3}} & 0 \\
 0 & \partial_{\Delta t} p_2/p_2 & 0 & 2\braket{\partial_{\Delta t} e_2|e_{4}} \\
 2\braket{ e_{3}|\partial_{\Delta t} e_1} & 0 & 0 & 0 \\
 0 & 2\braket{e_{4}|\partial_{\Delta t} e_2} & 0 & 0 \\
\end{array}
\right) \, , \\
L_{\Delta \omega} & =
\left(
\begin{array}{cccc}
\partial_{\Delta \omega} p_1/p_1 & 0 & 2i\braket{\partial_{\Delta \omega} e_1|e_{3}} & 0 \\
 0 & \partial_{\Delta \omega} p_2/p_2 & 0 &2i \braket{\partial_{\Delta \omega} e_2|e_{4}} \\
 -2i\braket{ e_{3}|\partial_{\Delta \omega} e_1} & 0 & 0 & 0 \\
 0 &-2i \braket{e_{4}|\partial_{\Delta \omega} e_2} & 0 & 0 \\
\end{array}
\right) \, ,
\end{align}
where
\begin{align}
\braket{\partial_{\Delta t} e_1|e_{3}} &= \braket{\partial_{\Delta \omega} e_1|e_{3}} =
\frac{1}{4} \left(e^{\frac{4 \Delta t^2 \sigma^4+\Delta \omega ^2}{8 \sigma^2}}+1\right)^{-1}
\sqrt{
4 \sigma^2 e^{\Delta t^2 \sigma^2+\frac{\Delta \omega ^2}{4 \sigma^2}}+e^{\frac{4 \Delta t^2 \sigma^4+\Delta \omega ^2}{8 \sigma^2}} \left(4 \Delta t^2 \sigma^4+\Delta \omega ^2\right)-4 \sigma^2
} 
\, , \nn
\braket{\partial_{\Delta t} e_2|e_{4}} & = \braket{\partial_{\Delta \omega} e_2|e_{4}} =
\frac{1}{4} 
\left(e^{\frac{4 \Delta t^2 \sigma^4+\Delta \omega ^2}{8 \sigma^2}}-1\right)^{-1}
\sqrt{
4 \sigma^2 e^{\Delta t^2 \sigma^2+\frac{\Delta \omega ^2}{4 \sigma^2}}
- e^{\frac{4 \Delta t^2 \sigma^4+\Delta \omega ^2}{8 \sigma^2}} \left(4 \Delta t^2 \sigma^4+\Delta \omega ^2\right)
- 4 \sigma^2
} \, .
\end{align}

We finally arrive at the QFI matrix: 
\begin{align}
H & =
\left(
\begin{array}{cc}
 H_{\Delta t^2} &  H_{\Delta t \Delta \omega} \\
 H_{\Delta t \Delta \omega} & H_{\Delta \omega^2} 
\end{array}
\right) \, .
\end{align}
where
\begin{align}
H_{\Delta t^2} & = \frac{1}{4 \sigma ^2} - \frac{\Delta \omega ^2}{4} \, \left( e^{\Delta t^2 \sigma^2 + \frac{\Delta \omega^2}{4 \sigma^2 }} - 1\right)^{-1} \, , \\
H_{\Delta \omega^2} & = \sigma ^2 - \frac{\Delta t^2}{4} 
\left( e^{\Delta t^2 \sigma^2 + \frac{\Delta \omega^2}{4 \sigma^2 }} - 1\right)^{-1}  \, , \\
H_{\Delta t \Delta \omega} & = \frac{\Delta t \Delta \omega }{4} \left( e^{\Delta t^2 \sigma^2 + \frac{\Delta \omega^2}{4 \sigma^2 }} - 1\right)^{-1} \, .
\end{align}

\noindent
Since
\begin{align}
    \text{Tr}(\rho [L_{\Delta t},L_{\Delta \omega}] ) =
    - i 
    + i 
    \left( e^{\Delta t^2 \sigma^2+\frac{\Delta \omega ^2}{4 \sigma^2}}-1 \right)^{-1} 
    \left( \Delta t^2 \sigma^2+\frac{\Delta \omega ^2}{4 \sigma^2} \right) \, .
\end{align}


\section{Relative range and velocity estimation with entangled photons} \label{sec:entangledsep}

We want to compare the single photon with entangled photon pairs (similar to quantum illumination \cite{lloyd2008enhanced,PhysRevLett.101.253601,PhysRevLett.114.080503}). For a fair comparison, we will send out one photon from an entangled photon pair, which has the same bandwidth. An entangled photon pair can be generated from an SPDC source.

If the signal photon is back-scattered, the two-photon state is described as
\begin{align}
\rho_\text{spdc} & = \frac{1}{2} \left(  \ket{\Psi_1}\bra{\Psi_1} +  \ket{\Psi_2}\bra{\Psi_2} \right) \, , 
\end{align}
where, in time domain, the two-photon wave functions are
\begin{align} \label{eq:spdc-time0}
\ket{\Psi_1} & = 
 \int d t \int d t_i ~ \Psi_1(t,t_i)  \ket{t} \ket{t_i} \, , \\
 \ket{\Psi_2} & = 
 \int d t \int d t_i ~ \Psi_2(t,t_i)  \ket{t} \ket{t_i} \, ,
\end{align}
and, for $j=1,2$,
\begin{align} 
\Psi_j(t,t_i) & = \sqrt{\frac{2}{\pi }} \sqrt[4]{1-\kappa ^2} \sqrt{\sigma \sigma_i} \, 
e^{- i \bar\omega_j (t - \bar t_j) - i \bar \omega_{i} (t_i - \bar t_i)} \nonumber \\
& \phantom{=}~ \times \exp \left[
- (t - \bar t_j)^2 \sigma^2 
- (t_i - \bar t_i)^2 \sigma_{i}^2 
+ 2 \kappa (t - \bar t) (t_i - \bar t_i) \sigma \sigma_{i} \right] \, .
\end{align}

The Hilbert space is spanned by the vectors $\ket{E_j}, j = 1,2,...,6$.
\begin{align}
\ket{E_1}   &= \frac{1}{\sqrt{2(1+|\delta'|)}}(\ket{\Psi_1} + e^{i\Delta t \Sigma\omega} \ket{\Psi_2}) \, , \\
\ket{E_2}  &= \frac{1}{\sqrt{2(1-|\delta'|)}}(\ket{\Psi_1} - e^{i\Delta t \Sigma\omega} \ket{\Psi_2}) \, , \\
\ket{E_3} &=  \frac{1}{\sqrt{c_{ 3}} }  \left( \ket{\partial_{\Delta t} E_1} - \braket{E_1|\partial_{\Delta t} E_1} \ket{E_1}\right) \, , \\
\ket{E_4} &=  \frac{1}{\sqrt{c_{ 4}} }  \left( \ket{\partial_{\Delta t} E_2} - \braket{E_2|\partial_{\Delta t} E_2} \ket{E_2} \right) \, . \\
\ket{E_{5}} &=  \frac{1}{\sqrt{c_{ 5}} } \left( \ket{\partial_{\Delta \omega} E_1} 
- \braket{E_1|\partial_{\Delta \omega} E_1} \ket{E_1}
- \braket{E_3|\partial_{\Delta \omega} E_1} \ket{E_3} \right)  \, , \\
\ket{E_{6}} &=  \frac{1}{\sqrt{c_{ 6}} }  \left(\ket{\partial_{\Delta \omega} E_2} 
- \braket{E_2|\partial_{\Delta \omega} E_2}\ket{E_2}
- \braket{E_4|\partial_{\Delta \omega} E_2}\ket{E_4} \right) \, .
\end{align}

The constants are
\begin{align}
c_3 & = \braket{\partial_{\Delta t} E_1|\partial_{\Delta t} E_1} - |\braket{\partial_{\Delta t} E_1|E_1}|^2 \, , \\
c_4 & = \braket{\partial_{\Delta t} E_2|\partial_{\Delta t} E_2} - |\braket{\partial_{\Delta t} E_2|E_2}|^2 \, , \\
c_5 & = \braket{\partial_{\Delta \omega} E_1|\partial_{\Delta \omega} E_1} - |\braket{E_1|\partial_{\Delta \omega}E_1}|^2
     - |\braket{E_3|\partial_{\Delta \omega}E_1}|^2 \, , \\
c_6 & = \braket{\partial_{\Delta \omega} E_2|\partial_{\Delta \omega} E_2} - |\braket{E_2|\partial_{\Delta \omega}E_2}|^2
     - |\braket{E_4|\partial_{\Delta \omega}E_2}|^2  \, .
\end{align}

The state can be diagonalised as
\begin{align}
\rho_\text{spdc} = & P_1  \ket{E_1}\bra{E_1} + P_2 \ket{E_2}\bra{E_2} \, , \\
P_1 =  &  \frac{1}{2}\left( 1+|\delta'| \right) \, , \quad P_2 = \frac{1}{2} \left( 1-|\delta'| \right) \, , \\
\delta' = & \langle \Psi_1 | \Psi_2 \rangle 
= e^{-  \frac{1}{2} \Delta t^2 \sigma_s^2  - \frac{\Delta \omega ^2}{8\left(1-\kappa ^2\right) \sigma_s^2} - i \Delta t \Sigma \omega } \, .
\end{align}


The SLD's for $\Delta t$ and $\Delta \omega$ are:
\begin{align}
L_{\Delta t} & =
\left(
\begin{array}{cccccc}
\partial_{\Delta t} P_1/P_1 & 0 & 2\braket{\partial_{\Delta t} E_1|E_3} & 0 &0 &0\\
 0 & \partial_{\Delta t} p_2/p_2 & 0 & \braket{\partial_{\Delta t} E_2|E_{4}}  &0 &0\\
 2\braket{ E_{3}|\partial_{\Delta t} E_1} & 0 & 0 & 0 &0 &0 \\
 0 & \braket{E_{4}|\partial_{\Delta t} E_2} & 0 & 0 &0 &0 \\ 
 0 & 0 & 0 & 0 & 0 & 0 \\
 0 & 0 & 0 & 0 & 0 & 0 \\
\end{array}
\right) \, , \\
L_{\Delta \omega} & =
\left(
\begin{array}{cccccc}
\partial_{\Delta \omega} P_1/P_1 & 0 & 2\braket{\partial_{\Delta \omega} E_1|E_3} & 0 &2\braket{\partial_{\Delta \omega} E_1|E_5} &0\\
 0 & \partial_{\Delta \omega} P_2/P_2 & 0 & 2\braket{\partial_{\Delta \omega} E_2|E_{4}}  &0 &2\braket{\partial_{\Delta \omega} E_2|E_6}\\
 2\braket{ E_{3}|\partial_{\Delta \omega} E_1} & 0 & 0 & 0 &0 &0 \\
 0 & 2\braket{E_{4}|\partial_{\Delta \omega} E_2} & 0 & 0 &0 &0 \\ 
 2\braket{ E_{5}|\partial_{\Delta \omega} E_1} & 0 & 0 & 0 & 0 & 0 \\
 0 & 2\braket{ E_{6}|\partial_{\Delta \omega} E_2} & 0 & 0 & 0 & 0 \\
\end{array}
\right) \, .
\end{align}

The terms in the SLD's are, for example:
\begin{align}
\braket{\partial_{\Delta \omega} E_1|E_3} & = 
\frac{1}{\sqrt{ c_3}}
\left(\braket{\partial_{\Delta \omega} E_1 |\partial_{\Delta t} E_1} - \braket{E_1|\partial_{\Delta t} E_1} \braket{\partial_{\Delta \omega} E_1 |E_1}\right) \, , \\
\braket{\partial_{\Delta \omega} E_1|E_5} &=
\frac{1}{\sqrt{c_5}}
\left( \braket{\partial_{\Delta \omega} E_1|\partial_{\Delta \omega} E_1} - \braket{E_1|\partial_{\Delta \omega} E_1} \braket{\partial_{\Delta \omega} E_1|E_1}
- \braket{E_3|\partial_{\Delta \omega} E_1} \braket{\partial_{\Delta \omega} E_1|E_3} \right)  \\
&= \sqrt{c_5} \, .
\end{align}

After some algebra, we obtain the following expression for the QFI matrix,
\begin{align}
H &=
\left(
\begin{array}{cc}
 Q_{\Delta t ^2} & Q_{\Delta t \Delta \omega} \\
 Q_{\Delta t \Delta\omega} & Q_{\Delta \omega^2} \\
\end{array}
\right) \, , 
\end{align}
with
\begin{align}
 Q_{\Delta t ^2} & = \sigma^2 
 - \frac{\Delta \omega ^2}{4} 
 \left(e^{\Delta t^2 \sigma^2 + \frac{\Delta \omega ^2}{4 \left(1-\kappa^2\right) \sigma^2}}-1\right)^{-1} \, , \\
  Q_{\Delta \omega^2} & = \frac{1}{1-\kappa^2} \frac{1}{4 \sigma^2}
  - \frac{\Delta t^2}{4} 
  \left(e^{\Delta t^2 \sigma^2 + \frac{\Delta \omega ^2}{4 \left(1-\kappa ^2\right) \sigma^2}}-1\right)^{-1} \, , \\
Q_{\Delta t \Delta\omega} & = \frac{\Delta t \Delta \omega }{4}
\left(e^{\Delta t^2 \sigma^2 + \frac{\Delta \omega ^2}{4 \left(1-\kappa^2 \right) \sigma^2}}-1\right)^{-1} \, .
\end{align}

The expectation value of the commutation relation then becomes
\begin{align}
    \text{Tr}(\rho [L_{\Delta t},L_{\Delta \omega}]) 
   = 
   i \left( e^{\Delta t^2 \sigma^2+\frac{\Delta \omega ^2}{4 \left(1-\kappa ^2\right) \sigma^2}} - 1  \right)^{-1}
   \left( \Delta t^2  \sigma^2+\frac{\Delta \omega ^2}{4\left(1-\kappa ^2\right) \sigma^2} \right)
   - i \, .
\end{align}

\section{A simultaneous measurement of time and frequency}\label{Sec:sim_measure}

In this Appendix, we present the construction by Zhuang \textit{et al.}\ \cite{PhysRevA.96.040304} for the optimal measurement of time and frequency estimation.
We consider a simplified model where the bandwidth of the signal photon remains constant. Consider the two-photon state given, in time domain, by Eq.\ (\ref{eq:njvhureowi}), where we put $\sigma = \sigma_{i0}$. To simplify the notation, we put 
$\bar t_{i0} = 0$ and $\bar \omega_{i0} = 0$.
The wave function factorises when written in terms of the variables $t_+ := t + t_i$ and $t_- := t - t_i$, 
\begin{align} 
\Psi(t_+ , t_-) 
\sim \exp{ \left[ - \frac{ (t_+ - \bar t)^2 (1-\kappa)\sigma^2}{2} - \frac{i \bar\omega (t_+ - \bar t)}{2} \right] } 
  \exp{ \left[ -  \frac{ ( t_- - \bar t)^2 (1+\kappa)\sigma^2}{2}
- \frac{i \bar\omega (t_- - \bar t)}{2}\right]} \, .
\end{align}

We can Fourier-transform the variable $t_+$, and express the wave function in terms of $\omega_+ = \omega + \omega_i$ and $t_-$, 
\begin{align} 
 \tilde \Psi(\omega_+,\tau_-) 
\sim \exp{ \left[ - \frac{ ( 2 \omega_+ - \bar\omega)^2}{8(1-\kappa)\sigma^2} + i \omega_+ \bar t \right] }  \exp{ \left[ -  \frac{ ( t_- - \bar t)^2 (1+\kappa)\sigma^2}{2}
- \frac{i \bar\omega (t_- - \bar t)}{2}\right]} \, .
\end{align}

This shows that we can jointly estimate $\bar\omega$ and $\bar t$ by first splitting the two photons, and then by applying intensity measurements in the variable $\omega_+$ and $t_-$. 
The probability density of measuring a photon at frequency $\omega_+$ and the other photon at time $t_-$ is
\begin{align} 
P(\omega_+,t_-) 
\sim \exp{ \left[ - \frac{ ( 2 \omega_+ - \bar\omega)^2}{4(1-\kappa)\sigma^2} \right] } 
\exp{ \left[ - ( t_- - \bar t)^2 (1+\kappa)\sigma^2 \right]} \, .
\end{align}

It follows that $\bar\omega$ and $\bar t$ can be estimated in this way with mean square errors
\begin{align}
    \delta t^2 = \frac{1}{2(1+\kappa)\sigma^2} \, , \qquad
    \delta \omega^2 = 2 (1-\kappa) \sigma^2 \, .
\end{align}
Putting this into the QCR bound, we obtain
\begin{align}
    \delta t^2 & = \frac{1}{2(1+\kappa)\sigma^2} \geqslant \frac{1}{4 \sigma^2} \, , \\
    \delta \omega^2 & = 2 (1-\kappa) \sigma^2 \geqslant (1-\kappa^2) \sigma^2\, .
\end{align}

In conclusions, this shows that this joint measurement is almost optimal for $\kappa \simeq 1$, and saturates the QCR bound in the limit that $\kappa \to 1$, i.e., infinite amount of entanglement.

\end{document}